\documentclass[a4paper]{article}
\usepackage[utf8]{inputenc}
\usepackage{amssymb}
\usepackage{amsfonts}
\usepackage{amsmath}
\usepackage{dsserif}
\usepackage{float}
\usepackage{bm}
\usepackage{amsmath}
\usepackage{makeidx}
\usepackage{amssymb}
\usepackage{graphicx}
\usepackage{float}
\usepackage[usenames]{color}
\begin{document}
\title{Semi-Implicit finite-difference methods to study 
the spin-orbit and coherently coupled spinor Bose-Einstein condensates}
\author{Paramjeet Banger\footnote{2018phz0003@iitrpr.ac.in},  
Pardeep Kaur\footnote{2018phz0004@iitrpr.ac.in},  
Sandeep Gautam\footnote{sandeep@iitrpr.ac.in}}
\maketitle


\begin{abstract}
We develop time-splitting finite difference methods, using 
implicit Backward-Euler and semi-implicit Crank-Nicolson discretization
schemes, to study the spin-orbit coupled spinor Bose Einstein condensates 
with coherent coupling in quasi-one and quasi-two-dimensional traps. 
The split equations involving kinetic
energy and spin-orbit coupling operators are solved using either time implicit 
Backward-Euler or semi-implicit Crank-Nicolson methods. We explicitly develop 
the method for pseudospin-1/2, spin-1 and spin-2 condensates. The results for ground states
obtained with time-splitting Backward-Euler and Crank-Nicolson methods are in 
excellent agreement with time-splitting Fourier spectral method which is one 
of the popular methods to solve the mean-field models for spin-orbit coupled 
spinor condensates.  We confirm the emergence 
of different phases in spin-orbit coupled pseudospin-1/2, spin-1 and spin-2 
condensates with coherent coupling.
\end{abstract}


\section{Introduction}
\label{Intro}
With experimental realization of optical traps \cite{stamper1998optical}, all the
hyperfine spin states of spin-$f$ ultracold bosonic atoms could be trapped and that
led to the discovery of $2f+1$ component Bose Einstein condensates (BECs) termed 
as spinor BECs \cite{stenger1998spin}. A spinor condensate can be described by a 
$2f + 1$ component order parameter that can vary over space and time 
\cite{kawaguchi2012spinor,stamper-rev}. Till date, spinor condensates in ultracold gases of 
spin-1/2 $^{87}$Rb \cite{myatt1997production}, spin-1 $^{23}$Na 
\cite{stamper1998optical}, spin-1 $^{87}$Rb \cite{chang2004observation}, spin-2 
$^{23}$Na \cite{gorlitz2003sodium}, spin-2 $^{87}$Rb \cite{chang2004observation} 
and spin-3 $^{52}$Cr atoms \cite{pasquiou2011spontaneous} have been experimentally 
realized. In later experiments \cite{soc-first}, spin-orbit coupling (SOC) was 
also engineered in neutral quantum gases like spinor BECs by controlling the atom light 
interaction that led to the generation of artificial gauge potentials coupled to 
the atoms \cite{non-abelian,non-abelian-2,review-goldman}. SOC was first 
realised experimentally in a BEC of $^{87}$Rb \cite{soc-first} by dressing two 
of its internal spin states from within the ground-state manifold by employing 
pair of Raman lasers that can create a momentum sensitive coupling between two 
internal atomic states resulting in an effective Zeeman shift. The strength of SOC can be 
tuned by Raman laser wavelength, whereas the coherent coupling can be tuned by the 
laser intensity \cite{spielman2013}. SOC and spin-dependent interactions provide 
a new platform to explore the novel phases in spin-orbit (SO) coupled spinor 
BECs \cite{wang-so-half-1,stringari-spin-half}. In the mean-field approximation, 
a spin-$f$ BEC in the presence of SO and coherent couplings can be well 
described by a set of $2f+1$ coupled nonlinear Gross-Pitaevskii equations (CGPEs) 
\cite{kawaguchi2012spinor,zhai2015degenerate,ho1998spinor,ciobanu2000phase}.
A wide range of numerical techniques have been employed in literature to study the scalar BEC
\cite{chang2005gauss,bao2005fourth,bao2004ground,ashhab2002external}
and spinor BECs \cite{wang2007time,bao2008computing,bao2013efficient, wang2014projection}. 
In our earlier works, we also provided sets of Fortran 90/95 codes to solve the mean-field
model of SO coupled $f = 1$ \cite{spin1-soc} and $f =2$ \cite{banger2020fortress}
spinor BECs with Rashba SO-coupling using time-splitting
Fourier spectral (TSFS) method. In the present work, we describe time-splitting finite-difference methods to solve the 
$2f+1$ CGPEs of spin-$f$ ($f= 1/2, 1, 2$) spinor BECs in quasi-one-dimensional (q1D),
quasi-two-dimensional (q2D) traps with SO and coherent couplings. 
The method can be easily extended to three-dimensional traps and higher spin system (say spin-3 BEC) if needed. We use the time-splitting 
Backward-Euler (TSBE) or time-splitting Crank-Nicolson (TSCN) finite-difference methods to solve the 
split equations corresponding to kinetic energy and spin-orbit coupling operators of spin-$f$ BEC.
These discretizaton schemes are employed with periodic boundary conditions and result in $2f+1$ decoupled sets of linear circulant systems of equations for each spatial dimension. 
The key property of a circulant matrix is that its columns (rows) can be written in terms of
powers of the shift matrix times the first column (row), which allows it to be diagonalized using the 
discrete Fourier transform \cite{rezghi2011diagonalization}.
The implementation of TSBE and TSCN is discussed in all its detail for an SO-coupled pseudospin-1/2
condensate, and then extended to higher spin condensates. The rest of this paper is
organized as follows: in section \ref{general_model}, we introduce 
a generic mean-field model suitable to describe the properties SO and
coherently coupled pseudospin-1/2, spin-1 and spin-2 BECs. In section \ref{numerical-method}, we discuss the TSBE and TSCN schemes to numerically solve the CGPEs, i.e. the mean-field model.
In section \ref{numerical-test}, we present the results for energies and component densities corresponding to the stationary states of these spinor BECs having $f$ = 1/2, 1 and 2. We also compare the results of the finite difference methods with the Fourier spectral method.

\section{Spinor condensates with spin-orbit and coherent coupling}
\label{general_model}
A generic spin$-f$ condensate with Rashba SO coupling can be modelled
at temperatures well below the critical temperature with a matrix
equation of form \cite{kawaguchi2012spinor,zhai2015degenerate}
\begin{equation}
\iota\frac{\partial \Psi}{\partial t} = 
\left(H_{\rm p} + H_{\rm coh}+H_{\rm d} + H_{\rm nd}\right)\Psi,\label{GPE}  
\end{equation}
where $\Psi$ is a $2f+1$ component order parameter, and $\iota = \sqrt{-1}$. In this work, we 
consider $f=1/2, 1, 2$ corresponding, respectively, to pseudospin-1/2, 
spin-1 and spin-2 condensates. In Eq.~(\ref{GPE}), $H_{\rm p}$ and 
$H_{\rm coh}$ are $2f+1\times2f+1$ matrix operators defined as
\begin{eqnarray}
H_{\rm p} &=& \mathbbb{1}\frac{\hat{p}_x^2+\hat{p}_y^2+\hat{p}_z^2}{2} + 
\gamma(S_x\hat{p}_y-S_y\hat{p}_x),\\
H_{\rm coh} &=& \frac{\Omega}{2}S_x,\label{h_coh}
\end{eqnarray}
where $\mathbbb{1}$ represents a $2f+1\times2f+1$ identity matrix, 
 $\gamma$ and $\Omega$ are the strengths of SO and coherent couplings,
 respectively, and $\hat{p}_\nu = -\iota\partial/\partial \nu$ with $\nu = x,y,z$.
$S_x$ and $S_y$ are the irreducible representations of the $x$ and $y$ components 
of angular momentum operators for spin-$f$ matrix, respectively.
The $(m',m)^{th}$ element of these $2f+1\times2f+1$ matrices are
\begin{eqnarray}
(S_x)_{m',m} = \frac{\zeta}{2}\left(\sqrt{(f(f+1)-m'm)}\delta_{m',m+1}
                 +\sqrt{(f(f+1)-m'm)}\delta_{m'+1,m}\right),\label{s_x}\\
(S_y)_{m',m} =  \frac{\zeta}{2i}\left(\sqrt{(f(f+1)-m'm)}\delta_{m',m+1}
            -\sqrt{(f(f+1)-m'm)}\delta_{m'+1,m}\right),\label{s_y}
\end{eqnarray}
here $m'$ and $m$ vary from $f$,$f-1,$ \ldots$, -f$. In Eqs.~(\ref{s_x})-(\ref{s_y}) $\zeta=2$ for $f=1/2$ and $\zeta=1$ for $f=1,2$. The interatomic interactions in the spinor condensate are accounted by diagonal 
matrix $H_{\rm d}$ and non-diagonal matrix $H_{\rm nd}$. The trapping
potential also enters into the $H_{\rm d}$ matrix. In the present work, we consider
the harmonic trapping potential for all the spinor condensates. These matrices for
a pseudospin-1/2 condensate are \cite{zhai2015degenerate}
\begin{eqnarray}
H_{\rm d} &= \begin{pmatrix}
V+\sum_{l=1}^2 g_{1l} |\psi_l|^2 & 0\\
0 & V + \sum_{l=1}^2 g_{2l} |\psi_l|^2 
\end{pmatrix},\quad H_{\rm nd} = 0,{\label{H_int_spin_1/2}}
\end{eqnarray}
where
\begin{equation} 
V = \frac{1}{2} \sum_\nu\alpha_\nu ^2 \nu^2, \nonumber\quad
g_{ll} = \frac{4\pi N a_{ll} }{a_{osc}}, \quad
g_{l,3-l} = \frac{4\pi Na_{l,3-l}}{a_{osc}}, \quad
\end{equation}
where $g_{ll}$ and $g_{l,3-l}$ with $l=1,2$ are intra- and inter-species interaction
strengths, respectively, $a_{\rm osc}$ is the oscillator length chosen as a unit of
length, $N$ is the total number of particles in the condensate, 
$\alpha_\nu = \omega_{\nu}/\omega_x$ with $\nu = x,y,z$ is
the ratio of confining-potential frequencies along $\nu$th direction 
to $x$ direction. The intraspecies interaction strengths are defined 
in terms of $s$-wave scattering lengths, $a_{11}$ and $a_{22}$, 
whereas interspecies interaction strength is defined in terms of
interspecies $s$-wave scattering length $a_{12} = a_{21}$. 
Similarly, these matrices for spin-1 condensate are \cite{kawaguchi2012spinor,ho1998spinor,machida-spin1}
\begin{subequations}
\begin{align}
H_{\rm d} =&
\begin{pmatrix}
{V}+{c_0}\rho+{c_1}(\rho_0+\rho_{-}) & 0 &0 \\
0 & {V}+{c_0}\rho +{c_1}\rho_{+}&0 \\
0 & 0 & {V}+{c_0}\rho+{c_1}(\rho_0-\rho_{-})
\end{pmatrix}{\label{h_d_spin1}},\\
H_{\rm nd} =&c_1
\begin{pmatrix}
0 & \psi_0\psi_{-1}^*&0 \\
\psi_0^*\psi_{-1} & 0 &\psi_0^*\psi_{1} \\
0 & \psi_0\psi_1^* & 0
\end{pmatrix}{\label{h_nd_spin1}},
\end{align}
\end{subequations}
here $\rho_l = |\psi_l|^2$ with $l=0,\pm1$, $\rho = \sum_l \rho_l$, and 
$\rho_{\pm} = \rho_{+1}\pm \rho_{-1}$ and
\begin{equation}
c_0 =
\frac{4\pi N(a_0+2a_2)}{3{a_{\rm osc}}},\quad
c_1 = \frac{4\pi N(a_2-a_0)}{3{a_{\rm osc}}}.
\end{equation}
The interaction strengths $c_0$ and $c_1$ are defined in terms
of $s$-wave scattering lengths $a_0$ and $a_2$. The subscript $0$ or 
$2$ in the scattering length characterises the total spin of the allowed
scattering channel. Lastly, $H_{\rm d}$ and $H_{\rm nd}$ for a spin-2 condensate
are \cite{kawaguchi2012spinor,ciobanu2000phase}
\begin{subequations}
\begin{align}
H_{\rm d}&=\text{diag}\left( 
h_{+ 2},h_{+1},h_0,h_{-1},h_{-2}\right),\label{h_d_spin2}\\
H_{\rm nd} &= 
\begin{pmatrix}
0 & h_{12} &h_{13}  & 0 &0\\
h_{12}^* &0 &h_{23}  & 0 &0\\
h_{13}^* & h_{23}^* & 0 & h_{34} &h_{35} \\
0 &0  & h_{34}^* &0 &h_{45} \\
0 &0  &h_{35}^* & h_{45}^*  & 0
\end{pmatrix}\label{h_nd_spin2},
\end{align}
\end{subequations} 
where
\begin{align*}
h_{\pm 2} &= V  + c_0 {\rho} \pm 2c_1 F_{z} + \frac{2}{5} 
c_2|\psi_{\mp2}|^2,\quad 
h_{0} = V  + c_0 {\rho} + \frac{1}{5}c_2 |\psi_{0}|^2,\\ 
h_{\pm 1} &= V  + c_0 {\rho} \pm c_1 F_{z} + \frac{2}{5} c_2 |\psi_{\mp1}|^2,\quad
h_{12} =  c_1 F_{-} - \frac{2}{5}c_2\psi_{-1} \psi_{-2}^*,\\  
h_{13} &= \frac{1}{5}c_2\psi_{0} \psi_{-2}^*,\quad 
h_{23} =  \frac{\sqrt{6}}{2}c_1 F_{-}- \frac{1}{5}
c_2\psi_{0} \psi_{-1}^*,\\
h_{34} &= \frac{\sqrt{6}}{2}c_1 F_{-}- \frac{1}{5}c_2\psi_{1} 
\psi_{0}^*,\quad
h_{35} = \frac{1}{5}c_2\psi_{2} \psi_{0}^*,\quad
h_{45} =  c_1 F_{-} - \frac{2}{5} c_2\psi_{2} \psi_{1}^*,
\end{align*}
and
\begin{subequations}
\begin{align}
F_z = & \sum_{l=-2}^2 l|\psi_l|^2,\quad
F_- = F_+^* = 2\psi_{-2}^* \psi_{-1} + \sqrt{6}\psi_{-1}^* 
\psi_0 + \sqrt{6} \psi_0^* \psi_1 + 2 \psi_2 \psi_1^*,\\
c_0 =& \frac{4\pi N(4 a_2 +3 a_4)}
{7 a_{\rm osc}},\,
c_1 =\frac{4\pi N(a_4-a_2)}
{7 a_{\rm osc}},\,
c_2=\frac{4\pi N(7a_0-10a_2+3a_4 )}
{7 a_{\rm osc}}.\label{c_i_spin2}
\end{align}
\end{subequations}
In Eq.~(\ref{c_i_spin2}), $c_0, c_1,$ and $c_2$ are three interaction
parameters, and $a_0, a_2, a_4$ are the $s$-wave scattering lengths
in the permitted scattering channels.

The order parameter for three spin systems is normalized to unity as
\begin{equation}
 \int \sum_{l}|\psi_l({\bf x},t)|^2 d {\bf x}= \sum_{l}{{\cal{N}}_l} = 1.
\end{equation}
The order parameter's norm along with the energy of these SO coupled spinor 
condensate, which is defined as
\begin{equation}
E = \int \left[\sum_{l,m}\psi_l^*\left(H_{\rm p} + H_{\rm coh} + H_{\rm d} + 
H_{nd}\right)_{lm}\psi_m\right]d{\bf x},    
\end{equation}
where $l,m$ run over species' labels, are the two conserved quantities for 
an SO-coupled condensate. In the present work, the species' labels are $1,2$ for
pseudospin-1/2, $1,0,-1$ for spin-1 and $2,1,0,-1,-2$ for spin-2 BECs. The species' labels $1$ and $2$ for pseudospin-1/2 BEC are equivalents of labels $1/2$ and $-1/2$, respectively, used in this work. For the sake 
of the compactness of the notations, the explicit functional dependence of $V$ on
${\bf x}$ and $\psi_l$ on ${\bf x}$ and $t$ has been suppressed. 
\section{Time-splitting Finite difference methods}
\label{numerical-method}
We describe the (semi)-implicit finite-difference schemes to numerically 
solve the coupled Gross–Pitaevskii equations (CGPEs) for 
SO-coupled spinor condensates. We use time-splitting Backward-Euler (TSBE)
and time-splitting Crank-Nicolson (TSCN) methods to solve the coupled sets of non-linear 
partial differential equations describing SO-coupled pseudospin-1/2, 
spin-1 and spin-2 BECs. The implementation is explained in all its 
detail for an SO-coupled pseudospin-1/2 condensate, and then extended 
to higher spin condensates. The results obtained with these finite 
difference schemes are compared with results from Fourier spectral method. 
The latter method has been used by us to solve CGPEs for SO-coupled spin-1
\cite{spin1-soc} and spin-2 condensates \cite{banger2020fortress}.  
\subsection{SO-coupled Pseudospin-1/2 Condensate}
\label{pseudospin-1/2}
\subsubsection{Quasi-one-dimensional pseudospin-1/2 BEC}
\label{Q1D_spin1/2}
We consider a two-component pseudospin-1/2 BEC confined by a harmonic
trapping potential with Rashba SO and coherent couplings. We first 
elaborate the method for solving one-dimensional CGPEs which describe an 
SO-coupled pseudospin-1/2 BEC trapped by a q1D trapping
potential. In such a trap, the $y$ and $z$ coordinates can be integrated out 
and after a rotation by $\pi/2$ about $z$-axis in spin-space which changes
$S_y$ to $-S_x$, the resultant matrix operator $H_{\rm p}$ is
\begin{equation}
H_{\rm p} = \mathbbb{1}\frac{\hat{p}_x^2}{2} - \gamma S_y\hat{p}_x
 \equiv \mathbbb{1}\frac{\hat{p}_x^2}{2} + \gamma S_x\hat{p}_x,\label{h_p_q1d}
\end{equation}
where $\mathbbb{1}$ is a $2\times2$ identity matrix, and $S_x$ and $S_y$
are Pauli spin matrices.
The form of $H_{\rm coh}$, $H_{\rm d}$, and $H_{\rm nd}$ remain same as 
in Eqs. (\ref{h_coh}) and
(\ref{H_int_spin_1/2}) with the caveat that 
\begin{equation} 
{\bf x} = x,\quad V = \frac{1}{2} \alpha_x^2 x^2, \nonumber\quad
g_{ll} = \frac{2 N a_{ll} \sqrt{\alpha_y \alpha_z}}{a_{osc}}, \quad
g_{l,3-l} = \frac{2Na_{l,3-l}\sqrt{\alpha_y \alpha_z}}{a_{osc}},
\end{equation}
where the terms have the same meanings as described in the previous section.
The time evolution of an SO-coupled spinor condensate
as per Eq.~(\ref{GPE}) is approximated by a first order operator splitting, wherein 
one is required to solve the 
following equations successively over the same period
\begin{subequations}
\begin{eqnarray}
\iota\frac{\partial \Psi}{\partial t} & = & H_{\rm p} \Psi\label{se1-1/2},\\
\iota\frac{\partial \Psi}{\partial t} & = & H_{\rm coh} 
\Psi\label{se2-1/2},\\
\iota\frac{\partial \Psi}{\partial t} & = & H_{\rm d}\Psi\label{se3-1/2},
\end{eqnarray}
\end{subequations}
where $\Psi({\bf x},t) = [\psi_1({\bf x},t),\psi_2({\bf x},t)]^T$ with $T$ denoting the transpose.
The matrix Eq. (\ref{se1-1/2}) in terms of coupled component 
equations is
\begin{subequations}
\begin{align}
 \iota \frac{\partial \psi_l(x,t)}{\partial t} =
-\frac{1}{2} \frac{\partial^2 \psi_l (x,t)}{\partial x^2}  - \iota \gamma 
\frac{\partial \psi_{3-l}(x,t)}{\partial x} ,\label{hp-spin-1/2}
\end{align}
\end{subequations}
where $l =1, 2$ is species' label. The spatial domain $x\in[-L_{x}/2,L_{x}/2)$ is
discretized via $N_x$ uniformly spaced points with a spacing of
$\Delta x$. The resulting one-dimensional space grid is $x_i = -L_{x}/2 + 
(i-1)\Delta x$ where $i = 1, 2,\ldots,N_{x}$. 
Using $\Delta t$ as the time-step to discretize time, the discrete
analogue of $\psi_l(x,t)$ is $\phi^n_{(i,l)}$ which represents
the value of $l$th component of the order parameter at spatial coordinate 
$x_i$ at time $n\Delta t$. The discretizaton scheme employs the periodic 
boundary conditions by ensuring that
\begin{equation}
\phi^n_{(1,l)} = \phi^n_{(N_x+1,l)},\quad\phi^n_{(0,l)} = \phi^n_{(N_x,l)}.  \label{pbc}
\end{equation}
In the present work, indices $l$ and $m$ are exclusively used for species' labels,
indices $i$ and $j$ are used to denote only space-grid point, $n$ is the index used for time, 
and $\nu = x, y, z$.
The discrete analogue of Eq.~(\ref{hp-spin-1/2}) using Backward-Euler  or Crank-Nicolson discretization schemes is 
\begin{align}
 \phi_{({i},l)}^{n+1} - \phi_{({i},l)}^{n} & =
\frac{\iota \Delta t}{4\Delta x^2} \left[\alpha \left(\phi_{({i+1},l)}^{n+1} -2 
\phi_{(i,l)}^{n+1} + \phi_{({i-1},l)}^{n+1}\right) + \beta \left(\phi_{({i+1},l)}^{n} -
2 \phi_{(i,l)}^{n} \right.\right.\nonumber\\ &\left.\left.+ \phi_{({i-1},l)}^{n}\right)\right] 
-\frac{ \gamma  {\Delta t}}{4\Delta x} \left[\alpha\left(\phi_{({i+1},3-l)}^{n+1}  - 
\phi_{({i-1},3-l)}^{n+1}\right) +\beta \left( \phi_{({i+1},3-l)}^{n} \right.\right.\nonumber\\
&\left.\left.- \phi_{({i-1},{3-l})}^{n}\right)\right],
\label{discrete}
\end{align}
where
$\alpha = 2, \beta = 0$ for Backward-Euler discretization, and
$\alpha = \beta = 1$ for Crank-Nicolson discretization.
The local truncation error incurred in Backward-Euler and
Crank-Nicolson discretizations are, respectively, of
the order $O(\Delta x^2 + \Delta t)$ and $O(\Delta x^2 + \Delta t^2)$ \cite{Numerical-Analysis}.
Considering {\em Backward-Euler} discretization first, Eq.~(\ref{discrete}) is
\begin{align}
\iota \frac{\phi_{({i},l)}^{n+1} - \phi_{({i},l)}^{n}}{\Delta t} & =
- \frac{\phi_{({i+1},l)}^{n+1} -2 
\phi_{(i,l)}^{n+1} + \phi_{({i-1},l)}^{n+1}}{2\Delta x^2}
-\iota \gamma\frac{\phi_{({i+1},3-l)}^{n+1}  - 
\phi_{({i-1},3-l)}^{n+1}}{2\Delta x}.\nonumber
\end{align}
For $l=1,2$, the time evolution as per Backward-Euler is
equivalent to
\begin{equation}
\begin{bmatrix}
\phi^{n+1}_{(i,1)}&\\
\phi^{n+1}_{(i,2)}
\end{bmatrix} = \left(\mathbbb{1}+\iota H_p\Delta t\right)^{-1}\begin{bmatrix}
\phi^n_{(i,1)}&\\
\phi^n_{(i,2)}
\end{bmatrix},\label{tebe}
\end{equation}
where 
\begin{equation}
    H_p \begin{bmatrix}
\phi^{n+1}_{(i,1)}&\\
\phi^{n+1}_{(i,2)}
\end{bmatrix} = \begin{bmatrix}
- \frac{\phi_{({i+1},1)}^{n+1} -2 
\phi_{(i,1)}^{n+1} + \phi_{({i-1},1)}^{n+1}}{2\Delta x^2}
-\iota \gamma_x\frac{\phi_{({i+1},2)}^{n+1}  - 
\phi_{({i-1},2)}^{n+1}}{2\Delta x}\\
- \frac{\phi_{({i+1},2)}^{n+1} -2 
\phi_{(i,2)}^{n+1} + \phi_{({i-1},2)}^{n+1}}{2\Delta x^2}
-\iota \gamma_x\frac{\phi_{({i+1},1)}^{n+1}  - 
\phi_{({i-1},1)}^{n+1}}{2\Delta x}
\end{bmatrix}.
\end{equation}
As $H_p$ is an Hermitian operator, time evolution operator
$(\mathbbb{1} +\iota H_p \Delta t)^{-1}$ in Backward-Euler discretization is not unitary leading to the norm being not
conserved. In contrast to this, the time evolution as per Crank-Nicolson is equivalent to
\begin{equation}
\begin{bmatrix}
\phi^{n+1}_{(i,1)}&\\
\phi^{n+1}_{(i,2)}
\end{bmatrix} = \frac{\mathbbb{1}-\iota H_p\Delta t}{\mathbbb{1}+\iota H_p\Delta t}\begin{bmatrix}
\phi^n_{(i,1)}&\\
\phi^n_{(i,2)}
\end{bmatrix},\label{tebe2}
\end{equation}
corresponding to a unitary operator $(\mathbbb{1}-\iota H_p\Delta t)/(\mathbbb{1}+\iota H_p \Delta t)$. The Backward-Euler method is therefore not suitable for realtime evolution in contrast to Crank-Nicolson method. 
Nonetheless, in imaginary time evolution, a non-unitary time evolution, used to obtain 
the stationary state solutions both Backward-Euler or Crank-Nicolson methods can be used.
Rewriting Eq.~(\ref{discrete}) as
\begin{align}
-&\frac{\iota \alpha \Delta t}{4 \Delta x^2}\left[\phi_{({i-1},l)}^{n+1}+\phi_{({i+1},l)}^{n+1}\right]
+ \left(1+ \frac {\iota \alpha \Delta t}{2 \Delta x^2}\right)\phi_{({i},l)}^{n+1} + 
\frac{\gamma \alpha \Delta t }{4 \Delta x}\left({\phi_{(i+1,3-l)}^{n+1} - \phi_{(i-1,3-l)}^{n+1}}\right)\nonumber\\ 
& =  \frac{\iota \beta  \Delta t}{4 \Delta x^2} \left[\phi_{(i-1,l)}^n + \phi_{(i+1,l)}^n\right]
+ \left(1-\frac{\iota \beta  \Delta t}{2 {\Delta x}^2}\right) \phi_{(i,l)}^n 
-\frac{\gamma \beta  \Delta t}{4 \Delta x }\left({\phi_{(i+1,3-l)}^n - \phi_{(i-1,3-l)}^n}\right). 
\label{discrete-gen}
\end{align}
Using Eq.~(\ref{pbc}) in Eq.~(\ref{discrete-gen}) with $i = 1,2,\ldots,N_x$ and $l = 1, 2$, the
resulting set of  2$N_x$ coupled linear algebraic equations can be written in matrix form as
\begin{equation}
 A \Phi_l^{n+1}+ B \Phi_{3-l}^{n+1} = D_l,\label{mat_eq}
\end{equation}
where A, B are circulant $N_x\times N_x$ matrices and $\Phi_l^{n+1}$, $D_l$ 
are $N_x \times 1$ matrices. These matrices can be expressed as
\begin{subequations}
\begin{align}
A(i,:) &= \left(
 1+\frac{\iota \alpha \Delta t}{ {2\Delta x}^2},
 ~ -\frac{\iota \alpha \Delta t}{4  {\Delta x}^2},~ 0,
 ~\cdots,~  0,~-\frac{\iota \alpha \Delta t}{4 {\Delta x}^2}\right)(C^{i-1})^T,\\
B(i,:) &= \left(
0,~ \frac{\alpha\Delta t \gamma}{4 \Delta x},~0,~\cdots,~ 0,
~ -\frac{\alpha\Delta t \gamma}{4 \Delta x} \right)(C^{i-1})^T,\\
\Phi_l^{n+1}  &= \begin{pmatrix}
\phi_{(1,l)}^{n+1}, &\phi_{(2,l)}^{n+1}, &\phi_{(3,l)}^{n+1},& \cdots& \phi_{(N_x,l)}^{n+1}\end{pmatrix}^T,\\
d_l(i) &= \left[
\frac{\iota \beta  \Delta t}{4 \Delta 
x^2}\left\{\phi_{(i-1,l)}^n+\phi_{(i+1,l)}^n\right\} +\left(1-\frac{\iota \beta  \Delta t}{2 {\Delta x}^2}\right)
\phi_{(i,l)}^n \right.\nonumber\\ &\left.-\frac{\gamma \beta  \Delta t}{4 \Delta x }\left({\phi_{(i+1,3-l)}^n
- \phi_{(i-1,3-l)}^n}\right) \right],
\end{align}
\label{discrete_matrix}
\end{subequations}
where $A(i,:)$ and $B(i,:)$ are the $i$th rows of A and B, respectively, $d_l(i)$ is the $i$th element of column
matrix $D_l$, and $C$ is defined as
\begin{equation}
C = \begin{bmatrix} 
    0 & 0 & \dots & 1\\
    1 & 0 & \ldots & 0\\
    \vdots & \ddots & &\vdots\\
    0 & \ldots & 1 & 0 
    \end{bmatrix}.
    \label{c-matrix}
    \end{equation}
For $l = 1, 2$, Eq.~(\ref{mat_eq}) represents two coupled matrix
equations which can be decoupled to yield
\begin{equation}
 (B^2 - A^2)\Phi_l^{n+1} = B D_{3-l} - A D_l,
\label{eqn_n+1}
\end{equation}
which for $l = 1 $ and $2$ represents two decoupled sets of 
linear circulant system of equations. Now, $B^2-A^2$ being
a circulant matrix, it can be diagonalised using Fourier 
matrix as \cite{rezghi2011diagonalization}

\begin{subequations}\label{cm_properties}
\begin{eqnarray}
    B^2-A^2 &=& F^{-1} \Lambda F,\quad \text{where}\\
    F_{i,j} &=& \frac{1}{\sqrt{N_x}}\exp{\left[-\frac{2\pi\iota} {N_x}(i-1)(j-1)\right]},\quad
   \text{and} \\
    \Lambda &=&\text{diag}[\sqrt{N_x} F \{B^2(:,1) -A^2(:,1)\}].
\end{eqnarray}
\end{subequations}

Now, the product of the Fourier matrix ($F$) with a one-dimensional 
array is equal to the discrete Fourier transform of the array,
and hence the solution to Eq. (\ref{eqn_n+1}) using  Eqs. (\ref{cm_properties}a)-(\ref{cm_properties}c) is \cite{rezghi2011diagonalization}
\begin{equation}
\Phi_l^{n+1} = {\rm IDFT} \left({\rm DFT}(B D_{3-l} - A D_l)./{\rm DFT}(B^2(:,1) -A^2(:,1))\right)\label{trans_sol},  
\end{equation}
where DFFT and IDFT stand for discrete forward Fourier and inverse discrete Fourier transforms, respectively, 
$A^2(:,1)$ and $B^2(:,1)$ denote the first columns of $A^2$ and $B^2$, and $./$ indicates the element wise division. 
Now, Eq. (\ref{se2-1/2}) is evolved in time from $t_n = n\Delta t$
to $t_{n+1} = (n+1)\Delta t$ considering Eq. (\ref{trans_sol}) as the solution
at $t_n$. The exact analytic solution to Eq. (\ref{se2-1/2})
is
\begin{equation} 
\Psi({\bf x},t_{n+1}) = \exp[-\iota H_{\rm coh}\Delta t]\Psi({\bf x},t_n) = \left[\mathbbb{1}\cos\left(\frac{\Omega\Delta t}{2}\right) 
-\iota S_x\sin\left(\frac{\Omega\Delta t}{2}\right)\right]\Psi({\bf x},t_n).
\label{sol_se2_1/2}
\end{equation}
The last step involves solving Eq.~(\ref{se3-1/2}) over the same period
treating the solution in Eq. (\ref{sol_se2_1/2}) as the solution at $t_n=n\Delta t$.
The exact solution to Eq. (\ref{se3-1/2}) is
\begin{equation} 
\Psi({\bf x},t_{n+1}) = \exp[-\iota H_{\rm d}\Delta t]\Psi({\bf x},t_n) \label{sol_se3_1/2}.
\end{equation}

\subsubsection*{Quasi-two-dimensional pseudospin-1/2 BEC}
\label{Q2D_spin1/2}
In a quasi-two-dimensional trap with tight confinement along $z$ axis, the form of matrix
operator $H_{\rm p}$ after integrating out the $z$ coordinate becomes  
\begin{equation}
H_{\rm p} = \mathbbb{1}\frac{\hat{p}_x^2+\hat{p}_y^2}{2} +\gamma(S_x\hat{p}_y-S_y\hat{p}_x),
\label{h_p_q2d}
\end{equation}
whereas the form $H_{\rm coh}$, $H_{\rm d}$, $H_{\rm nd}$ again remain unchanged from
those in Eqs.~(\ref{h_coh}) and (\ref{H_int_spin_1/2}) with a caveat that   
\begin{equation}
{\bf x} \equiv (x,y),\quad V = \frac{1}{2}(\alpha_x^2 x^2 +\alpha_y^2 y^2),\quad
g_{lm} = {\frac{2 N a_{lm} \sqrt{2 \pi \alpha_z}}{a_{osc}}}.
\end{equation}
 Using the time-splitting, the time evolution of the condensate from $t_n$ to $t_{n+1}$ 
 is approximated by successive solutions to the following equations
 over the same period
 \begin{subequations}
 \begin{eqnarray}
\iota \frac{\partial \Psi}{\partial t} & = & H_{p_x} \Psi\label{2dse1},\\
\iota \frac{\partial \Psi}{\partial t} & = & H_{p_y} \Psi\label{2dse2},\\
\iota \frac{\partial \Psi}{\partial t} & = & H_{\rm coh} \Psi\label{2dse3},\\
\iota \frac{\partial \Psi}{\partial t} & = & H_{\rm d}\Psi\label{2dse4},
\end{eqnarray}
\end{subequations}
where $H_{p_x}$ and $H_{p_y}$ are defined as 
\begin{equation}
H_{p_x} = \mathbbb{1}\frac{\hat{p}_x^2}{2} -\gamma S_y\hat{p}_x,\quad
H_{p_y} = \mathbbb{1}\frac{\hat{p}_y^2}{2}+\gamma S_x\hat{p}_y\label{kesoc-2dy}.
\end{equation}
Here, we consider a two-dimensional spatial grid 
defined as $\nu_i = -L_{\nu}/2 + (i-1)\Delta \nu$, 
where $i = 1,2, \ldots, N_{\nu}$, $\nu = x, y$, and $\Delta\nu$
is spatial step size. 
The discrete analogue of component
wavefunction is $\phi^n_{(i,j,l)}$ which is equal to value of the
$l$th wavefunction at space point $(x_i,y_j)$ at $t_n$ time. Similar to 
quasi-one-dimensional condensates, finite difference equivalents of each of 
Eq.~(\ref{2dse1}) and Eq.~(\ref{2dse2}) can be simplified to two decoupled matrix equations 
\begin{subequations}
\begin{eqnarray}
 ({B_x}^2 + {A_x}^2){X_{l}^{n+1}}& =& A_x D^x_l +(-1)^l B_x D^x_{3-l},\label{eqnx_n+1}  \\
 ({B_y}^2 - {A_y}^2){Y_{l}^{n+1}}& =& B_y D^y_{3-l} - A_y D^y_l,\label{eqny_n+1}
\end{eqnarray}
\end{subequations}
where $A_{\nu}$, $B_{\nu}$ (with $\nu = x, y$), $ X_l^{n+1}, Y_l^{n+1}, D^{\nu}_l$ 
are defined\\
 \begin{subequations}
\begin{align}
A_\nu(i,:) & = \left(
 1+\frac{\iota \alpha \Delta t}{ {2\Delta \nu}^2},~ -\frac{\iota \alpha \Delta t}{4  {\Delta\nu}^2},~ 0
,~\cdots,~  0,~-\frac{\iota  \alpha \Delta t}{4 {\Delta \nu}^2}\right)(C^{i-1})^T,\\
B_x(i,:) &= \left(
0,~ \frac{\iota  \alpha\Delta t \gamma}{4 \Delta x},~0,~\cdots,~ 0,-~\frac{\iota  \alpha\Delta t 
\gamma}{4 \Delta x} \right)(C^{i-1})^T,\\
B_y(i,:) &= \left(
0,~ \frac{ \alpha\Delta t \gamma}{4 \Delta y},~0,~\cdots,~ 0,~ -\frac{ \alpha\Delta t
\gamma}{4 \Delta y} \right)(C^{i-1})^T,\\
X_l^{n+1} & = \begin{pmatrix}
{\phi_{(1,j,l)}^{n+1}} &{\phi_{(2,j,l)}^{n+1}}&{\phi_{(3,j,l)}^{n+1}}&\cdots&
{\phi_{(N_x,j,l)}^{n+1}}
\end{pmatrix}^T,\\
Y_l^{n+1}  &= \begin{pmatrix}
\phi_{(i,1,l)}^{n+1}, &\phi_{(i,2,l)}^{n+1},&\phi_{(i,3,l)}^{n+1},& \cdots& \phi_{(i,N_y,l)}^{n+1}
\end{pmatrix}^T,\\
 d^x_l(i) &= \left[
\frac{\iota \beta  \Delta t}{4 \Delta 
x^2}\left\{\phi_{(i-1,j,l)}^n+\phi_{(i+1,j,l)}^n\right\} +\left(1-\frac{\iota \beta  \Delta t}{2 {\Delta x}^2}\right)
\phi_{(i,j,l)}^n \right.\nonumber\\ &\left.+\frac{(-1)^{l}\iota \gamma \beta  \Delta t}{4 \Delta x }\left({\phi_{(i+1,j,3-l)}^n
- \phi_{(i-1,j,3-l)}^n}\right) \right],\\
 d^y_l(i) &= \left[
\frac{\iota \beta  \Delta t}{4 \Delta 
y^2}\left\{\phi_{(i,j-1,l)}^n+\phi_{(i,j+1,l)}^n\right\} +\left(1-\frac{\iota \beta  \Delta t}{2 {\Delta y}^2}\right)
\phi_{(i,j,l)}^n \right.\nonumber\\ &\left.-\frac{\gamma \beta  \Delta t}{4 \Delta y }\left({\phi_{(i,j+1,3-l)}^n
- \phi_{(i,j-1,3-l)}^n}\right) \right],
\end{align}
\label{discrete_matx1}
\end{subequations}
where $A_\nu(i,:)$ and $B_\nu(i,:)$ are the $i$th row of $A_\nu$ and $B_\nu$, respectively, $d^\nu_l(i)$
is the $i$th element of column matrix $D^\nu_l$, and $C$ is defined in Eq. (\ref{c-matrix}).
For a fixed value of $j$ ($y$-index) and $l$ (species index), Eqs. (\ref{eqnx_n+1}) is a linear
circulant system of equations which can be solved by the same procedure as discussed to solve
Eq.~(\ref{eqn_n+1}). The solution to Eq. (\ref{2dse1}) is obtained by solving Eq. (\ref{eqnx_n+1})
for all $j$ and $l$ values following exactly the same procedure as
discussed Sec. \ref{Q1D_spin1/2}. This solution, then, is considered as an input solution at $t_n$ while
solving another set of linear circulant system of Eqs. (\ref{eqny_n+1}) over the same period from
$t_n$ to $t_n+\Delta t$. The solutions to Eqs. (\ref{2dse3})-(\ref{2dse4}) are again given as in
Eqs.~(\ref{sol_se2_1/2})-(\ref{sol_se3_1/2}) with $\Psi({\bf x},t_n)=[\psi_1{(x,y,t_n)},\psi_2{(x,y,t_n)}]^T$with $T$ standing for transpose.
 
\subsection{SO-coupled spin-1 condensate}
\label{spin1}
\subsubsection{Quasi-one-dimensional spin-1 BEC}
\label{Q1D_spin1}
In quasi-one-dimensional trap, $H_{\rm p}$ for an SO-coupled
spin-1 BEC takes the form
\begin{equation}
H_{\rm p} = \mathbbb{1}\frac{\hat{p}_x^2}{2} + \gamma S_x\hat{p}_x,
\end{equation}
where $\mathbbb{1}$ is a $3\times3$ identity matrix, and $S_x$ is the
$3\times 3$ spin-1 matrix. The form of $H_{\rm coh}$, $H_{\rm d}$, 
and $H_{\rm nd }$ in Eqs. (\ref{h_coh}), (\ref{h_d_spin1}), (\ref{h_nd_spin1}) remain
unchanged, provided 
\begin{equation*}
{\bf x} = x,\quad V = \frac{1}{2}\alpha_x^2 x^2,\quad
c_0 = \sqrt{\alpha_y \alpha_z} 
\frac{2N(a_0+2a_2)}{3{a_{\rm osc}}},\quad
c_1 = \sqrt{\alpha_y \alpha_z} \frac{2N(a_2-a_0)}{3{a_{\rm osc}}}.
\end{equation*}
Using the first order time-splitting, the solution of the Eq.~(\ref{GPE}) is equivalent 
to solving following equations successively
\begin{subequations}
\begin{eqnarray}
\iota\frac{\partial \Psi}{\partial t} &=& H_{\rm p}\Psi, \label{KESOC-s1}\\
\iota\frac{\partial \Psi}{\partial t} &=& \left(H_{\rm nd} + H_{\rm coh}\right)\Psi= 
H_{\rm nd+}\Psi, \label{SEpart}\\
\iota\frac{\partial \Psi}{\partial t} &=& H_{\rm d}\Psi. \label{SPpart}
\end{eqnarray}
\end{subequations}
where $H_{\rm nd+} = H_{\rm nd} + H_{\rm coh}$, and 
$\Psi({\bf x},t) = [\psi_{1}({\bf x},t),\psi_0({\bf x},t),\psi_{-1}({\bf x},t)]^T$.
We solve Eq.~(\ref{KESOC-s1}) using finite difference 
schemes described in detail for pseudospin-1/2 BEC.
Using Backward-Euler (and/or Crank-Nicolson) discretization schemes along with periodic 
boundary conditions, viz. Eq. (\ref{pbc}), Eq. (\ref{KESOC-s1}) reduces to three coupled 
matrix equations 
\begin{subequations}
\begin{eqnarray}
 A \Phi^{n+1}_{\pm1}+ B \Phi^{n+1}_{0}& =& D_{\pm1},\label{spin1-1}\\
 A \Phi^{n+1}_{0}+ B (\Phi^{n+1}_{1}+\Phi^{n+1}_{-1})& =& D_{0}.
 \label{spin1-2}
\end{eqnarray}
\end{subequations}
Eqs.~(\ref{spin1-1})-(\ref{spin1-2}), can be decoupled 
into following three independent matrix equations,
\begin{subequations}
\begin{eqnarray}
(2 B^2 A -A^3 ) \Phi_{\pm1}^{n+1}& =& (B^2-A^2) D_{\pm1}+ AB D_{0} -B^2D_{\mp1},\\
(A^2 -2B^2 ) \Phi_0^{n+1}&=&AD_0 -B(D_1+D_{-1}),
\end{eqnarray}
\label{decouple-spin1-1d}
\end{subequations}
 where $A$  and $\Phi_l^{n+1}$ with $l = 1, 0, -1$ are same 
 as in Eq.~(\ref{discrete_matrix}a) and Eq.~(\ref{discrete_matrix}c), respectively, 
 whereas rows of $B$ and elements of $D_j$ are now 
 defined as
\begin{subequations}
\begin{align}
B(i,:) &= \left(
0,~ \frac{\alpha\Delta t \gamma}{4 \sqrt{2}\Delta x},~0,~\cdots,~ 0,~ -\frac{\alpha\Delta t \gamma}{4 \sqrt{2} \Delta x} \right)(C^{i-1})^T\\
d_{\pm1}(i) &= \left[\frac{\iota \beta  \Delta t}{4 \Delta x^2}\left
\{\phi_{(i-1,{\pm1})}^n+\phi_{(i+1,{\pm1})}^n\right\} +\left(1-\frac{\iota \beta  \Delta t}
{2 {\Delta x}^2}\right) \phi_{(i,{\pm1})}^n \right.\nonumber\\
 &\left.-\frac{\gamma \beta  \Delta t}{4 \sqrt{2} \Delta x }\left({\phi_{(i+1,0)}^n - \phi_{(i-1,0)}^n}\right) \right]\\
 d_{0}(i) &= \left[
 \frac{\iota \beta  \Delta t}{4 \Delta x^2}\left\{\phi_{(i-1,{0})}^n+\phi_{(i+1,{0})}^n\right\} +
 \left(1-\frac{\iota \beta  \Delta t}{2 {\Delta x}^2}\right) \phi_{(i,{0})}^n \right.\nonumber\\
 &\left.-\frac{\gamma \beta  \Delta t}{4 \sqrt{2} \Delta x }\left({\phi_{(i+1,1)}^n - \phi_{(i-1,1)}^n}
 + {\phi_{(i+1,-1)}^n - \phi_{(i-1,-1)}^n}\right) 
 \right].
\end{align}
\label{discrete_matsp2}
\end{subequations}
The decoupled matrix Eqs. (\ref{decouple-spin1-1d}a) -(\ref{decouple-spin1-1d}b)
are linear circulant system of equations which can be solved by using the method
described for pseudospin-1/2 BEC. The analytic solution to
Eq.~(\ref{SEpart}) is \cite{spin1-soc}
\begin{equation}
\Psi({\bf x},t_{n+1}) \approx \left(\mathbbb{1} +\frac{\cos \zeta-1}{\zeta^2}\Delta t^2 H_{\rm nd+}^2-
\iota\frac{\sin{\zeta}}{\zeta}\Delta tH_{\rm nd+}\right)\Psi({\bf x},t_n),    
\end{equation}
where $\zeta = \Delta t \sqrt{|c_1 \psi_0\psi_{-1}^* + \frac{\Omega}{2\sqrt{2}}|^2 +|c_1\psi_0\psi_{1}^* +
\frac{\Omega}{2\sqrt{2}}|^2 }$.
Finally, the solution to Eq. (\ref{SPpart}) is again given as in Eq. (\ref{sol_se3_1/2}) with the caveat that the 
various quantities are identified as those corresponding
to spin-1 BEC.


\subsubsection*{Quasi-two-dimensional spin-1 BEC}
\label{Q2D_spin1}
Here the form of matrix operator $H_{\rm p}$ is same as in Eq. (\ref{h_p_q2d}) with $\mathbbb{1}$
representing a $3\times3$ identity matrix, and $S_\nu$ with $\nu = x, y$ denoting the spin-1 
matrices. Also, the form of $H_{\rm coh }$, $H_{\rm d}$, and $H_{\rm nd}$ 
in Eqs. (\ref{h_coh}), (\ref{h_d_spin1}),  (\ref{h_nd_spin1}), respectively, 
remain unchanged, provided
\begin{align}
{\bf x} &\equiv (x,y),\quad V =\sum_{\nu=x,y}\frac{\alpha_{\nu}^2\nu^2}{2},\\ 
c_0 &= \sqrt{2\pi\alpha_z}\frac{2 N(a_0+2a_2)}{3{a_{\rm osc}}},
\quad
c_1 = \sqrt{2\pi\alpha_z}\frac{2 N(a_2-a_0)}{3{a_{\rm osc}}}.
\label{interaction2d}
\end{align} 
The CGPEs of a quasi-2D spin-1 BEC with Rashba SO coupling can be split into following set 
of equations, and these has to be solved successively over the same period.
\begin{subequations}
\begin{eqnarray}
\iota\frac{\partial \Psi}{\partial t} &=& H_{p_x}\Psi, \label{spin-1x}\\
\iota\frac{\partial \Psi}{\partial t} &=& H_{ p_y}\Psi, \label{spin-1y}\\
\iota\frac{\partial \Psi}{\partial t} &=& \left(H_{\rm nd} + 
H_{\rm coh}\right)\Psi = H_{\rm nd+}\Psi, \label{se1-2d1}\\
\iota\frac{\partial \Psi}{\partial t} &=& H_{\rm d}\Psi. \label{se1-2d2}
\end{eqnarray}
\end{subequations}
where $H_{p_x}$ and $H_{p_y}$ are defined in Eq. (\ref{kesoc-2dy}) with $\mathbbb{1}$, and
$S_\nu$ being identified as $3\times3$ identity and spin-1 matrices, respectively.
Similar to quasi-two-dimensional pseudospin-1/2 BEC, each of Eq.~(\ref{spin-1x}) 
and Eq.~(\ref{spin-1y}) can be discretized  into three decoupled matrix 
equations, such as
\begin{subequations}
\begin{eqnarray}
(A^3_x +2A_xB^2_x) X_{\pm1}^{n+1}& =& (A^2_x+B^2_x) D^x_{\pm1} \mp A_xB_x D^x_{0} +B^2_x D^x_{\mp1},\\
(A^2_x +2B^2_x ) X_0^{n+1}&=&A_xD^x_0 + B_x(D^x_1 - D^x_{-1})
\end{eqnarray}
\label{discre-spin1_2dx}
\end{subequations}
for Eq.~(\ref{spin-1x}), and
\begin{subequations}
\begin{eqnarray}
(2 B^2_y A_y -A^3_y )Y_{\pm1}^{n+1}& =& (B^2_y-A^2_y) D_{\pm1}^y+ A_yB_y D^y_{0} -B^2_yD^y_{\mp1},\\
(A^2_y -2B^2_y ) Y_0^{n+1}&=&A_yD^y_0 - B_y(D^y_1+D^y_{-1}),
\end{eqnarray}
\label{discre-spin1_2dy}
\end{subequations}
for Eq.~(\ref{spin-1y}). Here, $A_{\nu}$ (with $\nu = x, y$),
$X_l^{n+1}$, $Y_l^{n+1}$,  are defined as in Eq.~(\ref{discrete_matx1}a),
Eq.~(\ref{discrete_matx1}d) and Eq.~(\ref{discrete_matx1}e) respectively, whereas $B_\nu , 
D^\nu_l$ are now defined as
\begin{subequations}
\begin{align}
B_x(i,:) &= \left(
0,~  \frac{\iota \alpha\Delta t \gamma}{4\sqrt{2}\Delta x},~0,~\cdots,~ 0,~  
-\frac{\iota \alpha\Delta t \gamma}{4\sqrt{2}\Delta x} \right)(C^{i-1})^T\\
B_y(i,:) &=\left(
0,~ \frac{\alpha\Delta t \gamma}{4 \sqrt{2}\Delta y},~0,~\cdots,~ 0,~ -\frac{\alpha\Delta t \gamma}
{4 \sqrt{2} \Delta y} \right)(C^{i-1})^T\\
 d^x_{\pm1}(i) &= \left[\frac{\iota \beta  \Delta t}{4 \Delta x^2}\left
\{\phi_{(i-1,j,{\pm1})}^n+\phi_{(i+1,j,{\pm1})}^n\right\} +\left(1-\frac{\iota \beta  \Delta t}
{2 {\Delta x}^2}\right) \phi_{(i,j,{\pm1})}^n \right.\nonumber\\
 &\left.\mp \frac{ \iota \gamma \beta  \Delta t}{4 \sqrt{2} \Delta x }\left({\phi_{(i+1,j,0)}^n - \phi_{(i-1,j,0)}^n}\right) \right]\\
 d^x_{0}(i) &= \left[ \frac{\iota \beta  \Delta t}{4 \Delta x^2}\left\{\phi_{(i-1,j,{0})}^n+
 \phi_{(i+1,j,{0})}^n\right\} +\left(1-\frac{\iota \beta  \Delta t}{2 {\Delta x}^2}\right)
 \phi_{(i,j,{0})}^n \right.\nonumber\\
 &\left.+\frac{\iota \beta  \Delta t \gamma}{4 \sqrt{2}\Delta x }\left({\phi_{(i+1,j,1)}^n -
 \phi_{(i-1,j,1)}^n} - ({\phi_{(i-1,j,-1)}^n - \phi_{(i+1,j,-1)}^n})\right) 
 \right].\\
 d^y_{\pm1}(i) &= \left[\frac{\iota \beta  \Delta t}{4 \Delta y^2}\left
\{\phi_{(i,j-1,{\pm1})}^n+\phi_{(i,j+1,{\pm1})}^n\right\} +\left(1-\frac{\iota \beta  \Delta t}
{2 {\Delta y}^2}\right) \phi_{(i,j,{\pm1})}^n \right.\nonumber\\
 &\left.-\frac{\gamma \beta  \Delta t}{4 \sqrt{2} \Delta y }\left({\phi_{(i,j+1,0)}^n - \phi_{(i,j-1,0)}^n}\right) \right]\\
 d^y_{0}(i) &= \left[
 \frac{\iota \beta  \Delta t}{4 \Delta y^2}\left\{\phi_{(i,j-1,{0})}^n+\phi_{(i,j+1,{0})}^n\right\} +
 \left(1-\frac{\iota \beta  \Delta t}{2 {\Delta y}^2}\right) \phi_{(i,j,{0})}^n \right.\nonumber\\
 &\left.-\frac{\gamma \beta  \Delta t}{4 \sqrt{2} \Delta y }\left({\phi_{(i,j+1,1)}^n - \phi_{(i,j-1,1)}^n}
 + {\phi_{(i,j+1,-1)}^n - \phi_{(i,j-1,-1)}^n}\right) 
 \right].
\end{align}
\label{discrete_matsp1}
\end{subequations}
Eqs. (\ref{discre-spin1_2dx}a)-(\ref{discre-spin1_2dx}b) and 
(\ref{discre-spin1_2dy}a)-(\ref{discre-spin1_2dy}b) are
linear circulant system of equations, and thus can be solved
as described for pseudospin-1/2 condensates in Sec. \ref{Q1D_spin1/2}. 
The solution to Eqs. (\ref{se1-2d1})-(\ref{se1-2d2}) is similar
as described for quasi-one-dimensional spin-1 condensates.
\subsection{SO-coupled spin-2 condensate}
\label{spin2}
\subsubsection{Quasi-one-dimensional spin-2 BEC}
Similar to quasi-one-dimensional pseudospin-1/2 and spin-1 BECs, form of $H_{\rm p}$ is 
$\mathbbb{1}\hat{p}_x^2/2 + \gamma S_x\hat{p}_x$ where $S_x$ denotes the spin-2 matrix and forms 
of $H_{\rm coh}$, $H_{\rm d}$ and 
$H_{\rm nd}$ remain the same as in Eqs.~(\ref{h_coh}), (\ref{h_d_spin2}), 
and  (\ref{h_nd_spin2}), respectively. The trapping potential and interaction parameters are
\begin{subequations}
\begin{align}
V({\bf x}) &= \frac{ \alpha_x^2x^2}{2},\quad
c_0 = \sqrt{\alpha_y \alpha_z}\frac{2 N(4 a_2 +3 a_4)}
{7 a_{\rm osc}},\\
c_1 &=\sqrt{\alpha_y \alpha_z}\frac{2 N(a_4-a_2)}
{7 a_{\rm osc}},\quad
c_2=\sqrt{\alpha_y \alpha_z}\frac{2 N(7a_0-10a_2+3a_4 )}
{7 a_{\rm osc}}
\end{align}
\end{subequations}
Using the first order time-splitting, the solution of the Eq.~(\ref{GPE}) is equivalent 
to solving following equations successively
\begin{subequations}
\begin{eqnarray}
\iota\frac{\partial \Psi}{\partial t} &=& H_{\rm p}\Psi, \label{KESOC-s2}\\
\iota\frac{\partial \Psi}{\partial t} &=& \left(H_{\rm nd} + H_{\rm coh}\right)
\Psi= H_{\rm nd+}\Psi, \label{SEpart-s2}\\
\iota\frac{\partial \Psi}{\partial t} &=& H_{\rm d}\Psi, \label{SPpart-s2}
\end{eqnarray}
\end{subequations}
where $H_{\rm nd+} = H_{\rm nd} + H_{\rm coh}$, and
$\Psi({\bf x},t) = [\psi_2({\bf x},t),\psi_1({\bf x},t),\psi_0({\bf x},t),\psi_{-1}({\bf x},t),\psi_{-2}({\bf x},t)]^T$.
Similar to pseudospin-1/2 and spin-1 condensates, finite difference  
discretization of Eq.~(\ref{KESOC-s2}) along with 
periodic boundary conditions, viz. Eq.~(\ref{pbc}), reduces it to five decoupled matrix equations
\begin{subequations}\label{ME_spin2_q1d}
\begin{align}
A(A^2-B^2)(A^3 -4AB^2)\Phi^{n+1}_{\pm2} =&  
-\sqrt{\frac{3}{2}}(B^4-A^2B^2)D_0 -\frac{3}{2}AB^3 D_{\mp1} \nonumber\\ & +\frac{3}{2} B^4 D_{\mp2} 
+  \left(\frac{5}{2}AB^3 - BA^3\right) D_{\pm1}\nonumber\\&
+\left(\frac{3}{2}B^4+A^4 -4A^2B^2\right)D_{\pm2}, \\
(A^2-B^2) (A^3 -4AB^2) \Phi^{n+1}_{\pm1}  =& 
     \sqrt{\frac{3}{2}}(B^3-A^2B)D_0 
    -\frac{3}{2}B^3D_{\mp2}+ \nonumber\\& \frac{3}{2}AB^2D_{\mp1}+\left( A^3 - \frac{5}{2}AB^2\right)
    D_{\pm1}\nonumber\\& +\left(\frac{5}{2}B^3 -A^2 B\right)D_{\pm2},\\
(A^3 -4AB^2) \Phi^{n+1}_0 =&(A^2 -B^2) D_0  - \sqrt{\frac{3}{2}} AB(D_1+D_{-1})\nonumber\\
    +&\sqrt{\frac{3}{2}} B^2(D_2+D_{-2}),
\end{align}
\end{subequations}
where $A$, $B$, and $\Phi^{n+1}_l$ are same as in  Eq.~(\ref{discrete_matrix}a),   
(\ref{discrete_matrix}b) and (\ref{discrete_matrix}c) respectively,
whereas the elements of column matrices $D_l$ with $l =2, 1, 0, -1, -2$ are
 \begin{subequations}
 \begin{align}
  d_{\pm2}(i) &= \left[\frac{\iota \beta  \Delta t}{4 \Delta 
               x^2}\left\{\phi_{(i-1,{\pm2})}^n+\phi_{(i+1,{\pm2})}^n\right\} 
               +\left(1-\frac{\iota \beta  \Delta t}{2 {\Delta x}^2}\right) 
               \phi_{(i,{\pm2})}^n \right.\nonumber\\
              &\left.-\frac{\gamma \beta  \Delta t}{4 \Delta x }\left({\phi_{(i+1,{\pm1})}^n 
              - \phi_{(i-1,{\pm1})}^n}\right) \right],\\
  d_{\pm1}(i) &= \left[\frac{\iota \beta  \Delta t}{4 \Delta                   
               x^2}\left\{\phi_{(i-1,{\pm1})}^n+\phi_{(i+1,{\pm1})}^n\right\} 
               +\left(1-\frac{\iota \beta  \Delta t}{2 {\Delta x}^2}\right) 
               \phi_{(i,{\pm1})}^n -\frac{\gamma \beta  \Delta t}{4 \Delta x }\right.\nonumber\\
              &\left.\times\left({\phi_{(i+1,{\pm2})}^n 
               - \phi_{(i-1,{\pm2})}^n} \right) -\sqrt{\frac{3}{2}} 
               \frac{\gamma \beta  \Delta t}{4 \Delta x } \left( {\phi_{(i+1,{0})}^n 
               - \phi_{(i-1,{0})}^n} \right) \right],\\
  d_{0}(i)   &= \left[\frac{\iota \beta  \Delta t}{4 \Delta 
              x^2}\left\{\phi_{(i-1,{0})}^n+\phi_{(i+1,{1})}^n\right\} +
              \left(1-\frac{\iota \beta  \Delta t}{2 {\Delta x}^2}\right) 
              \phi_{(i,{0})}^n 
              -\sqrt{\frac{3}{2}}\frac{\gamma \beta \Delta t}{4 \Delta x}\right.\nonumber\\
              &\left.\times
              \left({\phi_{(i+1,{1})}^n - \phi_{(i-1,{1})}^n} \right) -\sqrt{\frac{3}{2}} 
              \frac{\gamma \beta  \Delta t}{4 \Delta x } \left({\phi_{(i+1,{-1})}^n - 
              \phi_{(i-1,{-1})}^n} \right) \right].
 \end{align}
 \label{dicre-1d-sp2}
 \end{subequations}
The five decoupled sets of linear circulant system of 
Eqs.~(\ref{ME_spin2_q1d}a)-(\ref{ME_spin2_q1d}c) can be solved as discussed in Sec. 
\ref{Q1D_spin1/2}. The detailed procedure to solve Eq.~(\ref{SEpart-s2}) is discussed 
in the appendix, and the exact solution to Eq.~(\ref{SPpart-s2}) is same as in Eq.~(\ref{sol_se3_1/2}).
 \subsection{Quasi-two-dimensional spin-2 BEC}
 Here the form of matrix operator $H_{p_\nu}$ is same as in Eq. (\ref{h_p_q2d}) with $\mathbbb{1}$
representing a $5\times5$ identity matrix, $S_\nu$ are spin-2 matrices and the forms of
$H_{\rm coh }$, $H_{\rm d}$, and $H_{\rm nd}$ in Eqs.~(\ref{h_coh}), (\ref{h_d_spin2}),  
(\ref{h_nd_spin2}),
respectively, remain unchanged, with
\begin{align}
{\bf x} &\equiv (x,y),\quad V =\sum_{\nu=x,y}\frac{\alpha_{\nu}^2\nu^2}{2},c_0 = \sqrt{2\pi\alpha_z}
\frac{2 N(4 a_2 +3 a_4)}{7 a_{\rm osc}},\\
c_1 &= \sqrt{2\pi\alpha_z}\frac{2 N(a_4-a_2)}{7 a_{\rm osc}},\quad
c_2 = \sqrt{2\pi\alpha_z}\frac{2 N(7a_0-10a_2+3a_4 )}{7 a_{\rm osc}}.
\label{interaction2d+sp2}
\end{align} 
 Here also, similar to quasi-two-dimensional spin-1 BEC, each of Eq.~(\ref{spin-1x}) 
and (\ref{spin-1y}) can be discretized  into five decoupled matrix equations, such as
 \begin{subequations}
\begin{align}
 (4A_xB^2_x+A^3_x)(B^2_x+A^2_x) A_x X^{n+1}_{\pm2} & =  \sqrt{\frac{3}{2}}(B^4_x+A^2_xB^2_x )D^x_0 
 \mp\frac{3}{2}A_xB^3_x D^x_{\mp1}\nonumber\\ &+ \frac{3}{2} B^4_xD^x_{\mp2}  
 \mp (\frac{5}{2}A_xB^3_x + B_xA^3_x) 
 D^x_{\pm1} \nonumber\\ &+(\frac{3}{2}B^4_x+A^4_x +4A^2_xB^2_x)D^x_{\pm2},\\
(4A_xB^2_x+A^3_x)(B^2_x+A^2_x) X^{n+1}_{\pm1} & =  \mp \sqrt{\frac{3}{2}}(B^3_x+A^2_xB_x)D^x_0 \mp 
\frac{3}{2}B^3_xD^x_{\mp2}\nonumber\\ &  \pm (\frac{5}{2}B^3_x +A^2_xB_x)D^x_{\pm 2}  +
\frac{3}{2}A_xB^2_xD^x_{\mp1}\nonumber\\ &+( \frac{5}{2}A_xB^2_x + A^3_x)D^x_{\pm1},\\
(4A_xB^2_x+{A_x}^3) X^{n+1}_0 &= (B^2_x+A^2_x) D^x_0 + \sqrt{\frac{3}{2}} A_xB_x(D^x_{1}-D^x_{-1})\nonumber\\ 
& +\sqrt{\frac{3}{2}} B^2_x(D^x_2+D^x_{-2})
\end{align}
\label{discre-2dx-sp2}
\end{subequations}
for Eq.~(\ref{spin-1x}), and
\begin{subequations}
\begin{align}
 A_y(A^2_y-B^2_y)(A^3_y -4A_yB^2_y)Y^{n+1}_{\pm2} & =  -\sqrt{\frac{3}{2}}(B^4_y-A^2_yB^2_y)D^y_0 
 -\frac{3}{2}A_yB_y^3 D^y_{\mp1}\nonumber\\ &+ \frac{3}{2} B^4_y D^y_{\mp2} + (\frac{5}{2}A_yB^3_y - B_yA^3_y)
 D^y_{\pm1} \nonumber\\ &+(\frac{3}{2}B^4_y+A^4_y -4A^2_yB^2_y)D^y_{\pm2},\\
  (A^2_y-B^2_y) (A^3_y -4A_yB^2_y) Y^{n+1}_{\pm1} & = \sqrt{\frac{3}{2}}(B^3_y-A^2_yB_y)D^y_0  
  +\frac{3}{2}A_yB^2_yD^y_{\mp1} \nonumber\\ & -\frac{3}{2}B^3_yD^y_{\mp2} +( A^3_y - \frac{5}{2}A_yB^2_y)
  D^y_{\pm1}\nonumber\\ &+(\frac{5}{2}B^3_y -A^2_y B_y)D^y_{\pm2},\\
 (A^3_y -4A_yB^2_y) Y^{n+1}_0 &= (A^2_y -B^2_y) D^y_0  - \sqrt{\frac{3}{2}} A_yB_y(D^y_1+D^y_{-1})\nonumber\\
 & +\sqrt{\frac{3}{2}} B_y^2(D^y_2+D^y_{-2})
\end{align}
\label{discre-2dy-sp2}
\end{subequations}
for Eq.~(\ref{spin-1y}). $A_{\nu}$ (with $\nu = x,y$)$, X_l^{n+1}$, and $ Y_l^{n+1}$ are defined as in Eq. (\ref{discrete_matx1}a), Eq. (\ref{discrete_matx1}d) and
Eq. (\ref{discrete_matx1}e) respectively, and $B_\nu, D^\nu_l$  are now defined as
 \begin{subequations}
 \begin{align}
 B_{x}(i,:) &= \left(
0,~ \frac{ \iota \alpha\Delta t \gamma}{4 \Delta x},~0,~\cdots,~ 0,
~ -\frac{ \iota \alpha\Delta t \gamma}{4 \Delta x} \right)(C^{i-1})^T,\\
B_{y}(i,:) &= \left(
0,~ \frac{  \alpha\Delta t \gamma} {4 \Delta y},~0,~\cdots,~ 0,
~ -\frac{  \alpha\Delta t \gamma }{4 \Delta y} \right)(C^{i-1})^T,\\
  d^x_{\pm2}(i) &= \left[
 \frac{\iota \beta  \Delta t}{4 \Delta x^2}\left\{\phi_{(i-1,j,{\pm2})}^n+\phi_{(i+1,j,{\pm2})}^n\right\}
 +\left(1-\frac{\iota \beta  \Delta t}{2 {\Delta x}^2}\right) \phi_{(i,j,{\pm2})}^n \right.\nonumber\\
 &\left.\mp \frac{ \iota \gamma \beta  \Delta t}{4 \Delta x }\left({\phi_{(i+1,j,{\pm1})}^n - \phi_{(i-1,j,{\pm1})}^n}
 \right)  \right],\\
  d^x_{\pm1}(i) &= \left[\frac{\iota \beta  \Delta t}{4 \Delta x^2}\left\{\phi_{(i-1,j,{\pm1})}^n+
  \phi_{(i+1,j,{\pm1})}^n\right\} +\left(1-\frac{\iota \beta  \Delta t}{2 {\Delta x}^2}\right) 
  \phi_{(i,j,{\pm1})}^n \right.\nonumber\\ &\left.\pm\frac{ \iota\gamma \beta  \Delta t}{4 \Delta x }
  \left({\phi_{(i+1,j,{\pm2})}^n - \phi_{(i-1,j,{\pm2})}^n} \right) \mp \sqrt{\frac{3}{2}} 
  \frac{\iota \gamma \beta  \Delta t  }{4 \Delta x } \left( {\phi_{(i+1,j,{0})}^n - \phi_{(i-1,j,{0})}^n} \right) \right],\\
  d^x_{0}(i) &= \left[\frac{\iota \beta  \Delta t}{4 \Delta x^2}\left\{\phi_{(i-1,j,{0})}^n+\phi_{(i+1,j,{1})}^n\right\}
  +\left(1-\frac{\iota \beta  \Delta t}{2 {\Delta x}^2}\right) \phi_{(i,j,{0})}^n \right.\nonumber\\
 &\left.+\sqrt{\frac{3}{2}}\frac{\iota \gamma \beta \Delta t}{4 \Delta x }\left({\phi_{(i+1,j,{1})}^n -
 \phi_{(i-1,j,{1})}^n} \right) -\sqrt{\frac{3}{2}} \frac{ \iota \gamma \beta  \Delta t}{4 \Delta x } 
 \left({\phi_{(i+1,j,{-1})}^n - \phi_{(i-1,j,{-1})}^n} \right) \right],
 \end{align}
 \end{subequations}
 \begin{subequations}
 \begin{align}
 d^y_{\pm2}(i) &= \left[\frac{\iota \beta  \Delta t}{4 \Delta 
 y^2}\left\{\phi_{(i,j-1,{\pm2})}^n+\phi_{(i,j+1,{\pm2})}^n\right\} 
 +\left(1-\frac{\iota \beta  \Delta t}{2 {\Delta y}^2}\right)  \phi_{(i,{\pm2})}^n \right.\nonumber\\
&\left.-\frac{\gamma \beta  \Delta t}{4 \Delta y }\left({\phi_{(i,j+1,{\pm1})}^n 
 - \phi_{(i,j-1,{\pm1})}^n}\right) \right],\\
d^y_{\pm1}(i) &= \left[\frac{\iota \beta  \Delta t}{4 \Delta                y^2}\left\{\phi_{(i,j-1,{\pm1})}^n+\phi_{(i,j+1,{\pm1})}^n\right\} 
 +\left(1-\frac{\iota \beta  \Delta t}{2 {\Delta y}^2}\right) 
 \phi_{(i,j,{\pm1})}^n -\frac{\gamma \beta  \Delta t}{4 \Delta y }\right.\nonumber\\
 &\left.\times\left({\phi_{(i,j+1,{\pm2})}^n 
 - \phi_{(i,j-1,{\pm2})}^n} \right) -\sqrt{\frac{3}{2}} 
 \frac{\gamma \beta  \Delta t}{4 \Delta y } \left( {\phi_{(i,j+1,{0})}^n 
  - \phi_{(i,j-1,{0})}^n} \right) \right],\\
d^y_{0}(i)   &= \left[\frac{\iota \beta  \Delta t}{4 \Delta 
 y^2}\left\{\phi_{(i,j-1,{0})}^n+\phi_{(i,j+1,{1})}^n\right\} +
 \left(1-\frac{\iota \beta  \Delta t}{2 {\Delta y}^2}\right)   \phi_{(i,j,{0})}^n  -\sqrt{\frac{3}{2}}\frac{\gamma \beta \Delta t}{4 \Delta y}\right.\nonumber\\ &\left.\times
 \left({\phi_{(i,j+1,{1})}^n - \phi_{(i,j-1,{1})}^n} \right) -\sqrt{\frac{3}{2}} 
 \frac{\gamma \beta  \Delta t}{4 \Delta y } \left({\phi_{(i,j+1,{-1})}^n - 
 \phi_{(i,j-1,{-1})}^n} \right) \right].
 \end{align}
 \end{subequations}
where $B_{\nu}(i,:)$ is the $i$th row of $B_{\nu}$ and $d_{l}^{\nu}(i)$ is the $i$th element of column matrix $D_l^{\nu}$.
Eqs. (\ref{discre-2dx-sp2}a)-(\ref{discre-2dx-sp2}c) and 
(\ref{discre-2dy-sp2}a)-(\ref{discre-2dy-sp2}c) are
linear circulant system of equations, and thus can be solved in a similar manner
as described for pseudospin-1/2 and spin-1 condensates. 
The solution to Eqs. (\ref{se1-2d1})-(\ref{se1-2d2}) is on similar lines 
as described for quasi-one-dimensional spin-2 condensates.
 
\section{Numerical Results}
\label{numerical-test}
Here, we present the numerical results with TSBE and TSCN methods for the pseudospin-1/2, 
spin-1, and spin-2 in the presence as well as absence of coherent coupling. Both TSBE and TSCN can be used to obtain the ground state solutions of an SO and coherently coupled
spinor BEC. This can be achieved by considering an initial guess solution to the CGPEs 
and replacing $t$ by $-\iota t=\tilde{t}$ to solve CGPEs. The resultant imaginary time evolution is 
not norm preserving, and hence total norm needs to fixed to unity after each time iteration. 
The quantity $\tau = {\rm max}|\phi^{n+1}_{(i,j,l)} - \phi^{n}_{(i,j,l)}|/\Delta \tilde t$ serves 
as the convergence criterion to quantify convergence in imaginary time propagation. 
The stationary state solutions reported in this section has been obtained with 
$\tau=10^{-6}$. In contrast to imaginary time evolution, realtime dynamics of the spinor BECs can be studied with TSCN and not with TSBE as the later does not conserve 
norm as was discussed in Sec. \ref{Q1D_spin1/2}.

\subsection{Pseudospin-1/2 }
\begin{figure}[hbt!]
 \includegraphics[clip, trim=0.5cm 7cm 0.5cm 7cm, width=1.00\textwidth]{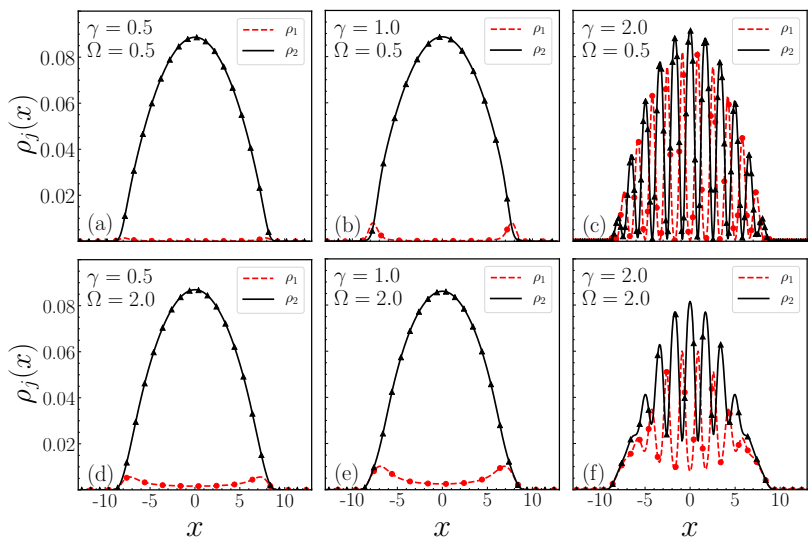}
\caption {(Color online) (a)-(c) are the component densities for an SO-coupled q1D $^{87}$Rb 
pseudospin-${\frac{1}{2}}$ BEC with $\Omega=0.5$ and $\gamma=0.5, 1, 2$, respectively. The same for $\Omega = 2$
are shown in (d)-(f), respectively. The lines and points
correspond to the results from TSBE and TSCN, respectively.
The interaction strengths considered in (a)-(f) are
$g_{22} = 0.9g_{11}$ and $g_{12} = 1.1g_{11}$ with $g_{11} =446.95$.}
\label{fig1}
\end{figure}

\begin{figure}[hbt!]
\includegraphics[width=1\linewidth]{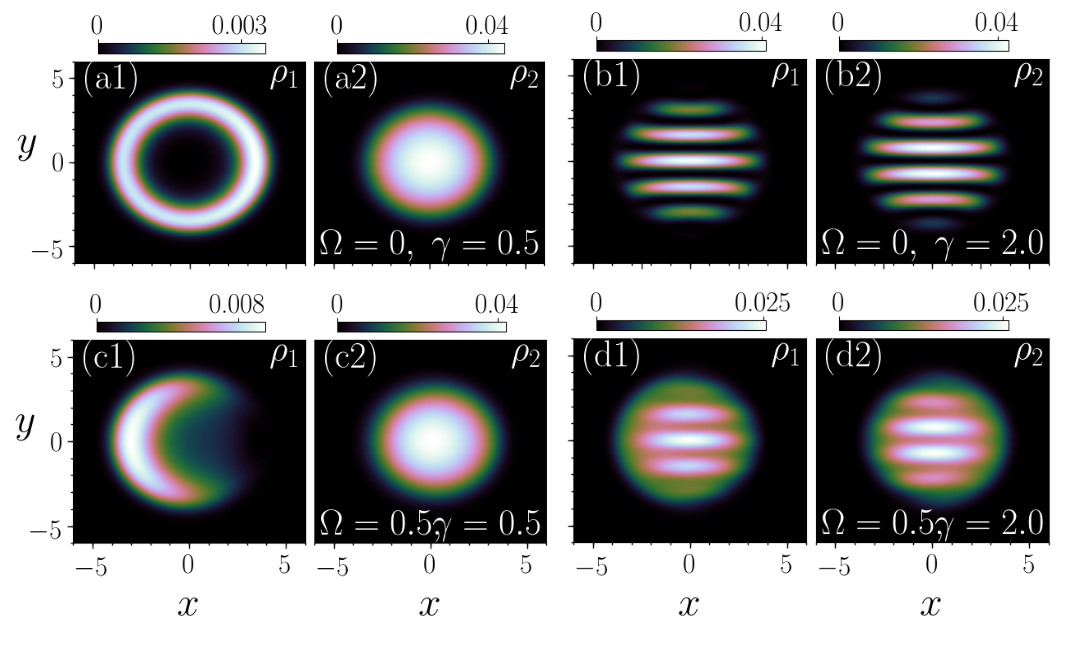}
\caption {(Color online) (a1)-(a2) are the component densities obtained with TSCN method for an SO-coupled q2D $^{87}$Rb pseudospin-${\frac{1}{2}}$ BEC with $g_{11} = 250,~g_{22}$ = 225, $g_{12} = g_{21}$ = 275, $\gamma=0.5$ and $\Omega=0$. The same for $(\gamma,\Omega) = (2,0), (0.5,0,5)$, and $(2,0.5)$ are shown in (b1)-(b2), (c1)-(c2), and (d1)-(d2), respectively.}
\label{fig2}
\end{figure}

\begin{figure}[hbt!]
\includegraphics[width=0.50\linewidth]{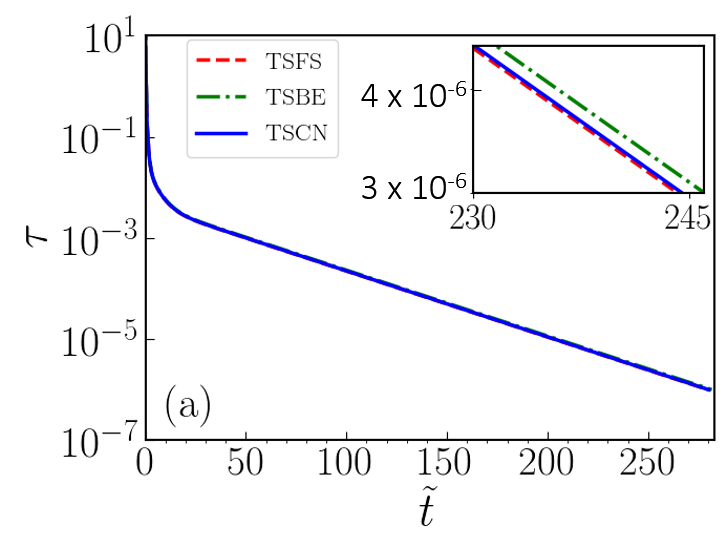}
\includegraphics[width=0.49\linewidth]{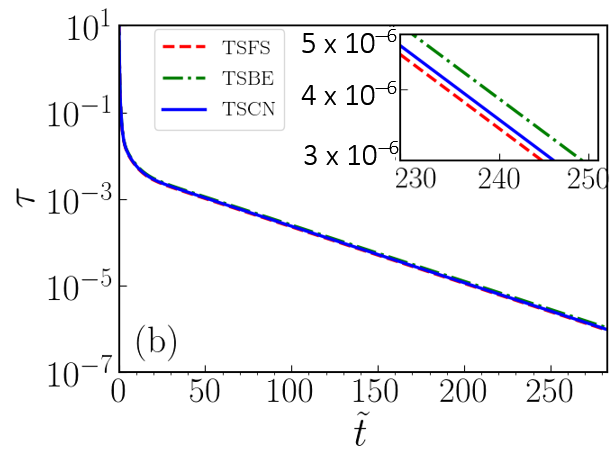}\\
\caption{(Color online) The variation of convergence criterion during imaginary-time propagation to calculate the ground state of q1D $^{87}$Rb. In (a), we have chosen $\Delta x =0.1$ and $\Delta \tilde{t}= 0.01$, whereas for (b) $\Delta x =0.2$ and $\Delta \tilde{t}= 0.02$.}
\label{error}
\end{figure}

\begin{figure}[H]
\includegraphics[width=0.325\linewidth]{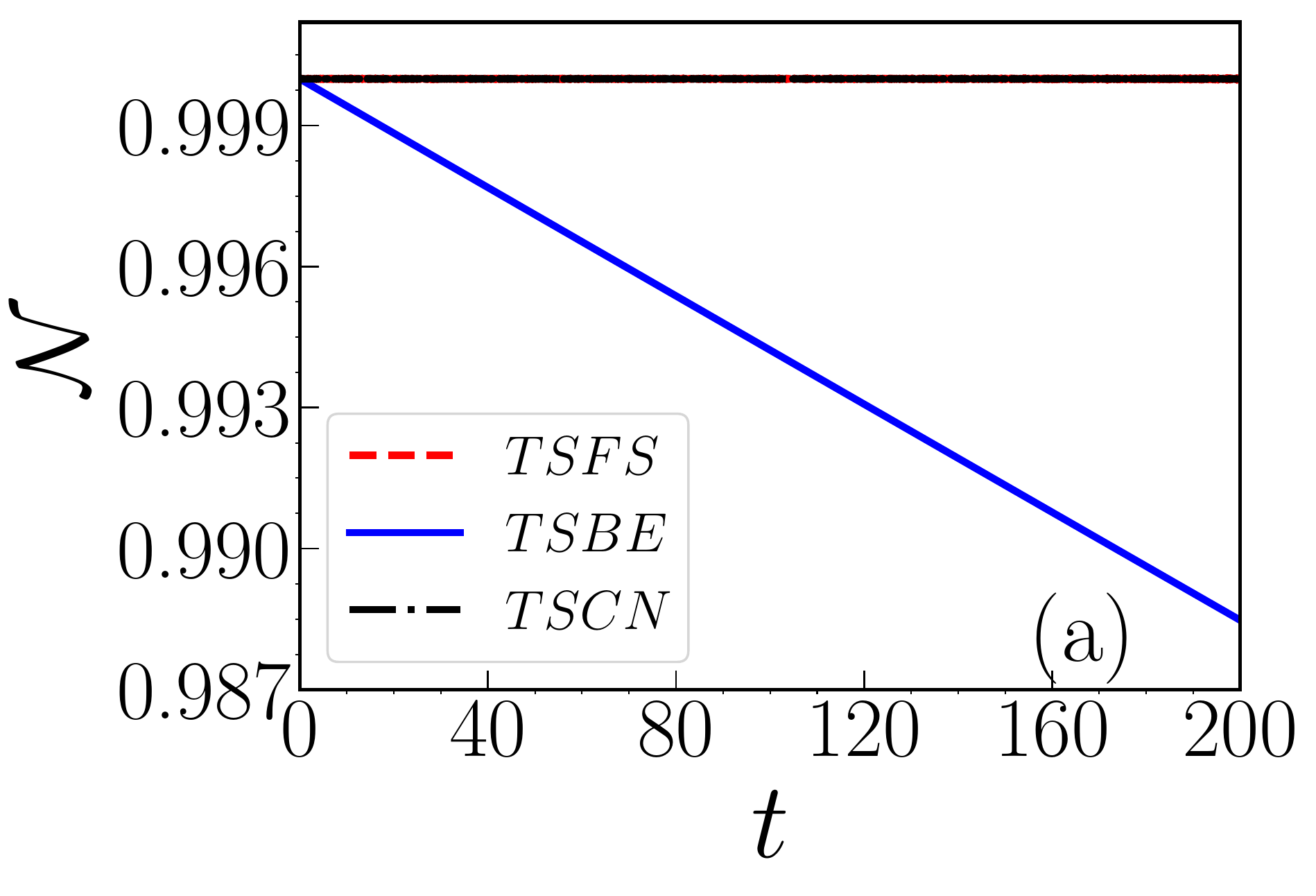}
\includegraphics[width=0.32\linewidth]{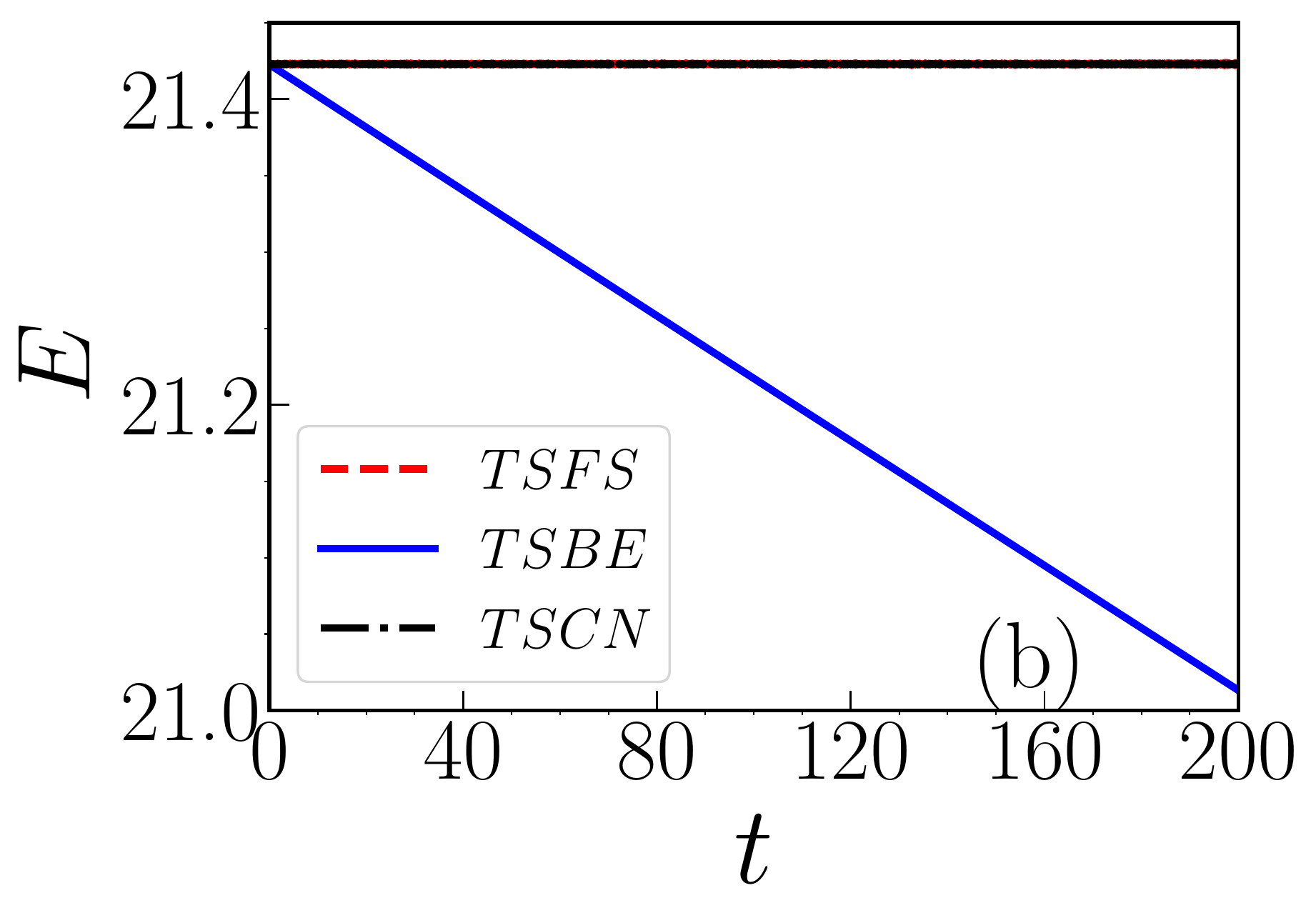}
\includegraphics[width=0.32\linewidth]{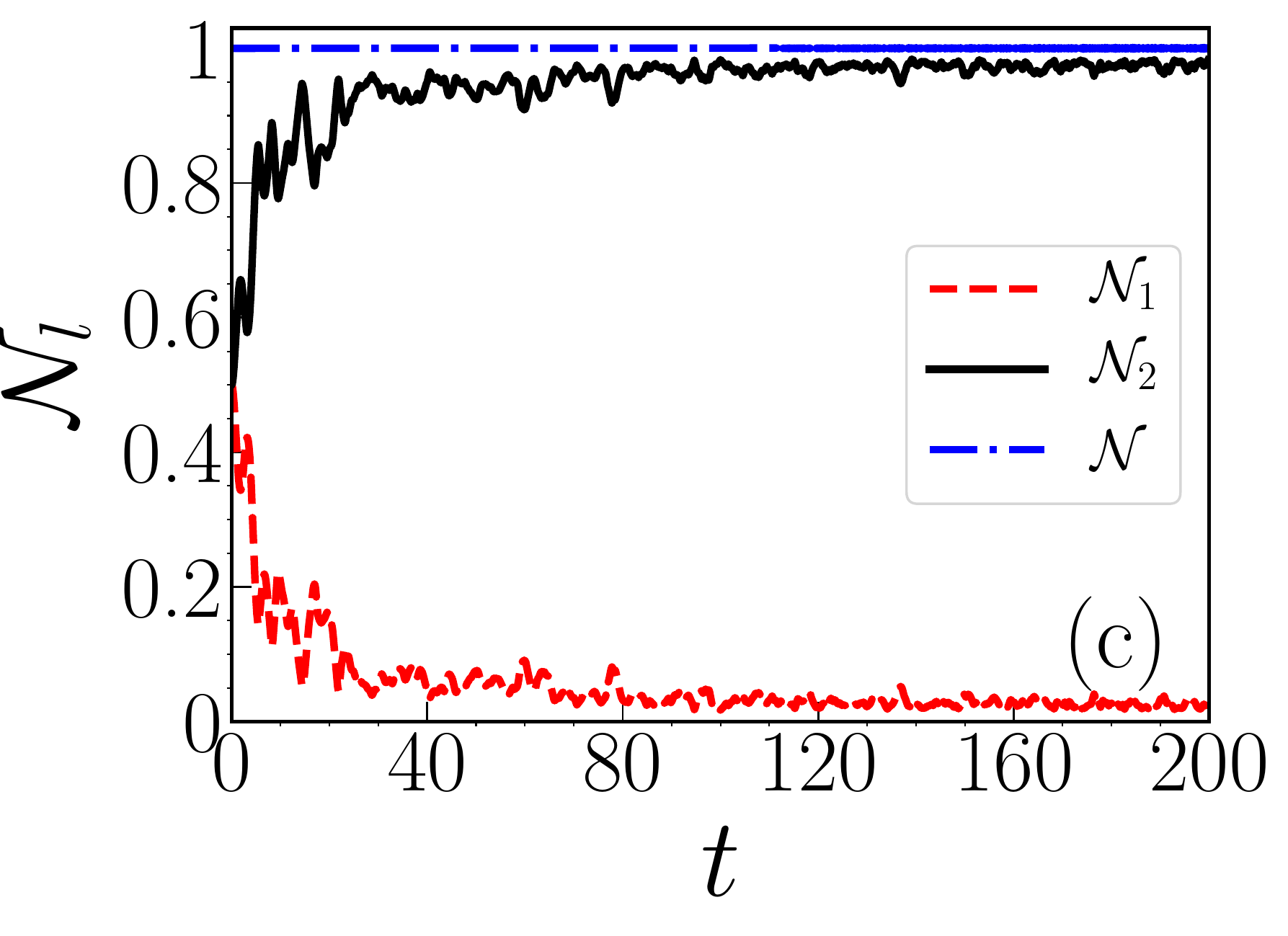}
\caption{(Color online) (a) Norm ${\cal N}$ as a function of time and (b) energy $E$ as a function of time for the
ground state solution of pseudospin-1/2 BEC of $^{87}$Rb
with $\gamma = \Omega = 0.5$. (c) Norm ${\cal N}$ and ${\cal{N}}_l$ as a function of time in realtime obtained for non-stationary initial solution using TSCN. The real-time evolution of initial solution is obtained using TSFS, TSBE and TSCN with $\Delta x =0.1$ and $\Delta{t}= 0.005$.}
\label{comparison}
\end{figure}

For pseudospin-1/2 case, we choose an experimentally realizable $^{87}$Rb
pseudospinor-1/2 BEC with scattering length $a_{11}=101.8a_B$,
interaction strengths $g_{12} = 1.1g_{11}$, $g_{22}
= 0.9g_{11}$ and $g_{12} = g_{21}$, where $a_B$ is the Bohr radius.
We consider 5000 atoms trapped in q1D trapping potential with $\omega_x = 2\pi\times 
20$Hz, $\omega_y = 2\pi\times 400$Hz and $\omega_z = 2\pi\times 400$Hz. 
The interaction strengths in dimensionless units are given as
\begin{equation}
(g_{11},~g_{22},~g_{12} ) =(446.95,~ 402.26,~491.65),\label{g_q1D}
\end{equation}
with $g_{12}=g_{21}$. For q2D BEC, we consider 
$5000$ atoms of  $^{87}$Rb in a trap with trapping frequencies $\omega_x = \omega_y = 2\pi\times 20$Hz, $\omega_z = 2\pi\times 400$Hz.
For this case, the interaction strengths  $g_{22}= 0.9g_{11}$, $g_{12} = 1.1g_{11}$, and 
$g_{12} = g_{21}$ for $a_{11} = 101.8a_B$ are given as
\begin{equation}
(g_{11},~g_{22},~g_{12}) =(250,~ 225,~275),\label{g_q2D}
\end{equation}
with $g_{12}=g_{21}$. In both these cases, we compare the results from TSFS, 
TSBE and TSCN in the presence as well as absence of coherent coupling and find an 
excellent agreement. The comparison of the ground state energies obtained with three methods for different 
values of $\gamma$
are given in Table-{\ref{table1}} for $\Omega = 0$ and Table-{\ref{table2}} 
for $\Omega=0.5$. The results with TSBE and TSCN are in very good agreement with those
from TSFS.
\begin{table}[H]
\caption{Comparison of ground state energies of pseudospin-1/2 BEC of $^{87}$Rb obtained with TSFS, TSBE
and TSCN for different values of $\gamma$ in
the absence of coherent coupling $\Omega$. The  interaction strength parameters are
 $g_{11} = 446.95$, $g_{22} = 402.26$ and $g_{12} = g_{21} = 491.65$ for q1D BEC, whereas the same for q2D BEC are $g_{11} = 250.52$, $g_{22} = 225.47$ and  $g_{12} = g_{21}= 275.57$.}
\begin{center}
\begin{tabular}{c c c c c c c c }
\hline
&&
\multicolumn{3}{c}{$\Delta x=0.1$,$\Delta\tilde t=0.01$} & \multicolumn{3}{c}{$\Delta x=0.1$,$\Delta\tilde t=0.005$} \\
  \hline
\multicolumn{1}{c}{}&
\multicolumn{1}{c}{$\gamma$}&
\multicolumn{1}{c}{TSFS}&
\multicolumn{1}{c}{TSBE}&
\multicolumn{1}{c}{TSCN}&
\multicolumn{1}{c}{TSFS}&
\multicolumn{1}{c}{TSBE}&
\multicolumn{1}{c}{TSCN}\\
\hline
q1D & 0.5& 21.4357& 21.4357& 21.4357&21.4357&21.4357&21.4357\\
& 1.0& 21.4186& 21.4186& 21.4186&21.4186&21.4186&21.4186\\
& 1.5& 21.3333& 21.3334& 21.3333&21.3324&21.3324&21.3324\\
& 2.0& 20.7018& 20.7035& 20.7022&20.7001&20.7014&20.7011\\
\hline
q2D & 0.5& 5.7201& 5.7201& 5.7201&5.7201&5.7201&5.7201\\
& 1.0&5.4707 & 5.4707& 5.4707&5.4707&5.4707&5.4707\\
& 1.5&4.8520 & 4.8520& 4.8520&4.8518&4.8518&4.8518\\
& 2.0&3.9783 & 3.9786& 3.9787&3.9883&3.9786&3.9786\\
\hline
\end{tabular}
\label{table1}
\end{center}
\end{table}
\begin{table}[hbt!]
\caption{Comparison of ground state energies of pseudospin-1/2 BEC obtained with TSFS, TSBE and TSCN  for $\Omega =0.5$ and different values of $\gamma$. The results have been obtained with $\Delta x = 0.1$ and $\Delta \tilde{t} = 0.01$. The interaction strength 
parameters considered for q1D BEC are $g_{11}=446.95$, $g_{22}=402.26$ and  $g_{12}= g_{21} =491.65$, whereas the
same for q2D BEC are $g_{11} = 250.52$, $g_{22} = 225.47$ and $g_{12} = g_{21} = 275.57$.}
\begin{center}
\begin{tabular}{c c c c c c}
\hline
\multicolumn{1}{c}{}&
\multicolumn{1}{c}{$\gamma$}&
\multicolumn{1}{c}{TSFS}&
\multicolumn{1}{c}{TSBE}&
\multicolumn{1}{c}{TSCN}\\
\hline
q1D & 0.5 & 21.4231& 21.4231& 21.4231\\
& 1.0  & 21.4002& 21.4002& 21.4002\\
& 1.5& 21.3033& 21.3034& 21.3033\\
& 2.0  & 20.6711& 20.6727& 20.6715\\
\hline
q2D & 0.5& 5.6457& 5.6457& 5.6457\\
& 1.0& 5.3339& 5.3339&  5.3339\\
& 1.5& 4.7181& 4.7181& 4.7181\\
& 2.0& 3.8434 &3.8438&3.8438\\
\hline
\end{tabular}
\label{table2}
\end{center}
\end{table}
The component densities corresponding to ground
state solutions obtained with TSBE and TSCN methods for q1D $^{87}$Rb BEC are shown in Fig.~\ref{fig1}. The densities obtained with two methods are in an excellent agreement.
Similarly, the component densities, obtained with TSCN 
method, for q2D $^{87}$Rb BEC
for different values of $\gamma$ and $\Omega$ are shown
in Fig.~\ref{fig2}. 
We also study the variation of the convergence criterion 
as a function of $\tilde t$ in imaginary-time propagation with TSBE, TSCN, and TSFS to obtain the ground state solution. As an example, in the imaginary-time propagation
to obtain the ground state of q1D pseudospin-1/2 BEC of $^{87}$Rb starting with normalized Gaussian initial guess
wavefunctions for the two components, the variation of $\tau$ as a function of $\tilde t$, obtained with three methods, is shown in Fig.~\ref{error}(a) for $\Delta x =0.1$ and $\Delta\tilde{t}= 0.01$ and in Fig.~\ref{error}(b)
for $\Delta x =0.2$ and $\Delta\tilde{t}= 0.02$. It is evident that TSCN shows faster convergence than TSBE.   
As discussed in the Sec. \ref{Q1D_spin1/2}, the TSBE does
not lead to a unitary time evolution in contrast to TSCN. 
In order to confirm this, we consider the real-time evolution of the ground state solution of the q1D $^{87}$Rb
shown in Fig. \ref{fig1}(a) with TSBE and TSCN. For this
we consider the ground state solution corresponding to interaction parameters in Eq. (\ref{g_q1D}) with $\gamma = \Omega =0.5$ as the initial solution at $t=0$ in real-time
evolution. The variation of total norm and energy as a function of time obtained using TSFS, TSBE, and TSCN are shown in 
Fig.~\ref{comparison}(a)-(b), respectively. The non-conservation of norm and hence energy in TSBE makes the method unsuitable to study
any realtime dynamics. The dynamics of the ground state,
a stationary state, is trivial in the sense that besides norm and energy the expectation values of various 
operators are also conserved. 
Next, we consider the dynamics of non-stationary state
using TSCN. We first obtain a non-stationary state by solving CGPEs for q1D $^{87}$Rb with interaction strengths
as defined in Eq.~(\ref{g_q1D}) and $\gamma = \Omega = 0.5$
under the constraint of zero polarization. The solution thus obtained is non-stationary, and is then evolved in
realtime (without any additional constraint) using TSCN.
The variation of component norms as a function of time is
shown in Fig.~\ref{comparison}(c).
\subsection{Spin-1}
We consider $(1)$ $^{23}$Na and $(2)$ $^{87}$Rb spin-1 BECs corresponding to antiferromagnetic and ferromagnetic phases in the absence of coupling. The scattering lengths  corresponding to system $(1)$ and $(2)$ are $a_0=50.00a_B$, $a_1=55.01 a_B$ \cite{scattering-Na-spin1} and $a_0=101.8a_B$, $a_1=100.4 a_B$ \cite{scattering-Rb-spin1}, respectively. We consider 10000 atoms trapped in q1D trapping potential with $\omega_x = 2\pi\times 20$Hz, $\omega_y = \omega_z = 2\pi\times 400$Hz. The interaction strengths $c_0$ and $c_2$ in dimensionless units are given as
\begin{subequations}
\begin{eqnarray}
{\rm (1)} \quad (c_0,c_2) &=& (240.83, 7.54),\\
{\rm (2)} \quad (c_0,c_2) &=& (885.72, -4.09).
\end{eqnarray}
\end{subequations}
The same number of atoms trapped in q2D trapping potential with $\omega_x = \omega_y =2\pi\times20$Hz,
$\omega_z = 2\pi\times 400$Hz leads to following interaction strengths
\begin{subequations}
\begin{eqnarray}
{\rm (1)} \quad (c_0,c_2) &=& (134.98, 4.22),\\
{\rm (2)} \quad (c_0,c_2) &=& (248.22, -1.15),
\end{eqnarray}
\end{subequations}
for $^{23}$Na and $^{87}$Rb spin-1 BECs, respectively.
The oscillator lengths for system (1) and (2) are $4.69~\mu$m and $2.41~\mu$m, respectively. For these two cases, the comparison of ground state energies obtained from TSFS, 
TSBE and TSCN shows an excellent agreement as reported in Table-(\ref{table3}).
\begin{table}[H]
\caption{Comparison of ground state energies of SO and coherently coupled spin-1 BECs using TSFS, TSBE, and TSCN methods with $\Delta x = 0.1$ and $\Delta {\tilde t} = 0.005$. The energies correspond to different values $\gamma$. The coherent coupling used for q1D and q2D systems are $0.5$ and $0.1$, respectively.}
\begin{center}
\begin{tabular}{c c c c c c c c}
\hline
\multicolumn{1}{c}{}&
\multicolumn{1}{c}{$\gamma$}&
\multicolumn{3}{c}{$^{23}$Na}&
\multicolumn{3}{c}{$^{83}$Rb}\\
&&TSFS&TSBE&TSCN&TSFS&TSBE&TSCN\\
\hline 
& 0.5& 15.0623&15.0623&15.0623&35.7812&35.7812&35.7812\\
q1D& 1.0&14.6873&14.6873&14.6873&35.4062&35.4062&35.4062\\
&1.5&14.0623&14.0623&14.0623&34.7812&34.7812&34.7812\\
&2.0&13.1873&13.1876&13.1876&34.9062&33.9062&33.9065\\
\hline
& 0.5& 4.3797&4.3797&4.3797&8.2638&8.2638&8.2638\\
q2D& 1.0&3.9602&3.9602&3.9601&7.8747&7.8747&7.8747\\
&1.5&3.3303&3.3303&3.3303&7.2435&7.2435&7.2435\\
&2.0&2.4486&2.4489&2.4489&6.3658 &6.3661&6.3661\\
\hline
\end{tabular}
\label{table3}
\end{center}
\end{table}

\begin{figure}[hbt!]
\includegraphics[clip, trim=0.5cm 7cm 0.5cm 7cm, width=1.00\textwidth]{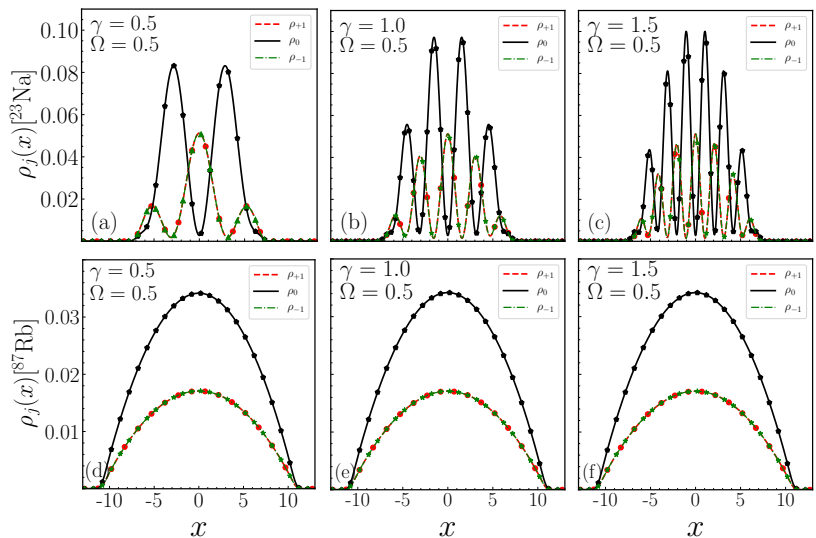}
\caption {(Color online) (a)-(c) are the ground state component densities for an SO-coupled q1D $^{23}$Na spin-1 BEC for different values $\gamma$ and $\Omega = 0.5$. The
same for $^{87}$Rb spin-1 BEC are shown in (d)-(f). In (a)-(f), lines and points correspond to the results from TSBE and TSCN, respectively.}
\label{figure2-sp1-1d}
\end{figure}

\begin{figure}[hbt!]
\begin{tabular}{ccc}
\includegraphics[width=10.0cm]{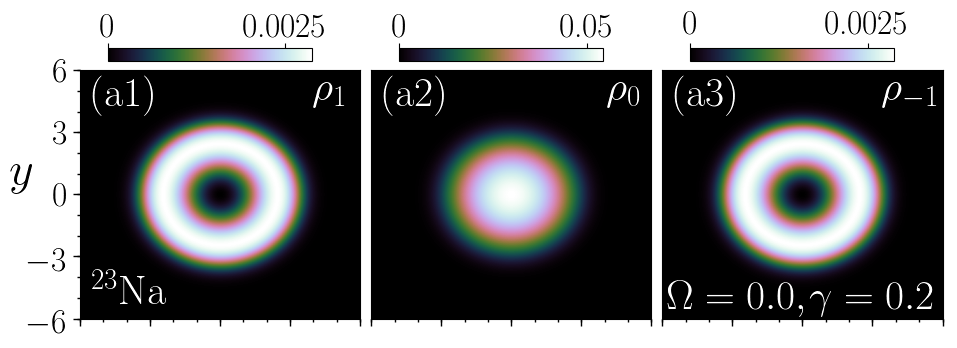}\\[-0.1in]
\includegraphics[width=10.0cm]{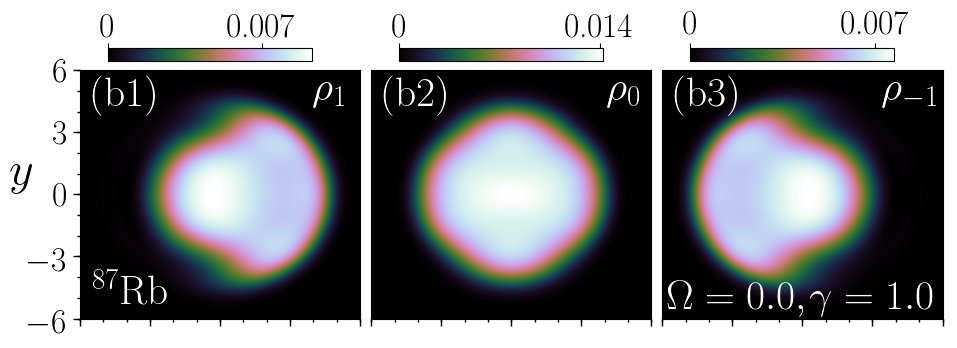}\\[-0.1in]
\includegraphics[width=10.0cm]{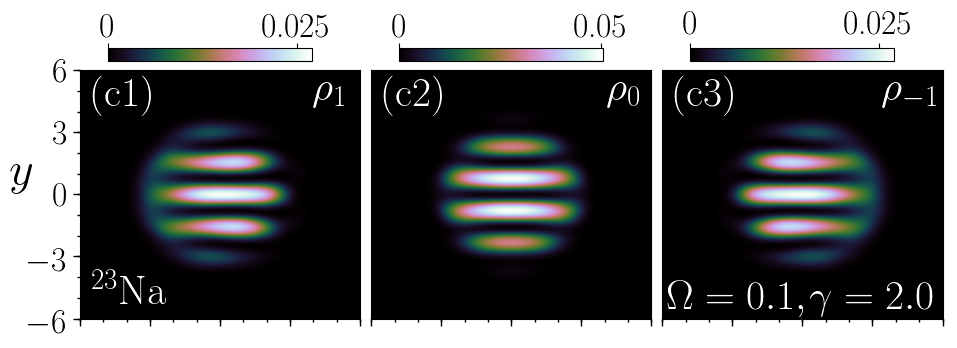}\\[-0.1in]
\includegraphics[width=10.05cm]{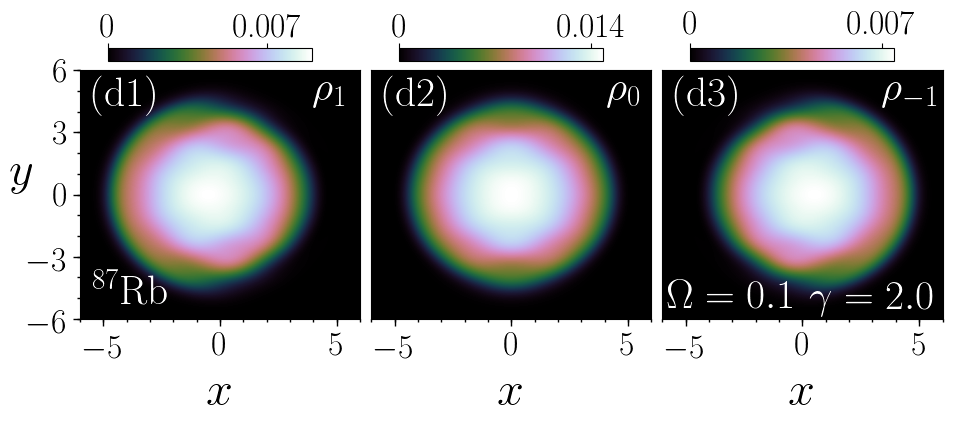}
\end{tabular}
\caption {(Color online) (a1)-(a3) are the ground-state component densities for an SO-coupled q2D $^{23}$Na spin-1 BEC with $\gamma=0.2$ and $\Omega = 0$, whereas the same for $^{87}$Rb are in (b1)-(b3). (c1)-(c3) and (d1)-(d3) are the ground-state component densities for $^{23}$Na and $^{87}$Rb BECs, respectively, with $\gamma = 2$ and $\Omega = 0.1$.}
\label{figure2-sp1-coh2d}
\end{figure}

The numerically obtained component densities in the ground states of harmonically trapped q1D $^{23}$Na 
and $^{87}$Rb spin-1 BECs with different values of $\gamma$ and $\Omega$ are shown
in Fig.~\ref{figure2-sp1-1d}. 
The component densities obtained using TSBE and TSCN are in an excellent agreement.
Similarly, in Fig.~(\ref{figure2-sp1-coh2d}) we have shown some distinct ground state density profiles 
for q2D $^{23}$Na and $^{87}$Rb spin-1 BECs obtained using TSCN.
\subsection{Spin-2}
We consider (1) $^{83}$Rb, (2) $^{23}$Na, and (3) $^{87}$Rb spin-2 BECs corresponding to ferromagnetic, anti-ferromagnetic
and cyclic phases. The three sets of scattering length corresponding to these systems are \cite{ciobanu2000phase,widera2006precision}
\begin{eqnarray}
{(1)}\quad a_0 &=& 83.0a_B,\quad a_2 = 82.0a_B,\quad a_4 = 81.0a_B;\\
{(2)}\quad a_0 &=& 34.9a_B,\quad a_2 = 45.8a_B,\quad a_4 = 64.5a_B;\\
{(3)}\quad a_0 &=& 87.93a_B,\quad a_2 = 91.28a_B,\quad a_4 = 99.18a_B,
\end{eqnarray}
respectively. We consider $10000$ atoms of each of these systems trapped in q1D trapping potential with $\omega_x
= 2\pi\times 20$Hz, $\omega_y = \omega_z = 2\pi\times 400$Hz. 
The interaction strengths $c_0$, $c_1$ and $c_2$ in dimensionless units are given as  
\begin{eqnarray*}
{(1)}\quad (c_0,c_1,c_2) &=& (699.62, -1.23, 4.90),\\
{(2)}\quad (c_0, c_1,c_2) &=& (242.97, 12.06, -13.03),\\
{ (3)}\quad (c_0,c_1,c_2) &=& (831.26, 9.91, 0.31).
\label{param}
\end{eqnarray*}
Similarly, we consider 10000 atoms of each of three systems trapped in a q2D trapping potential with $\omega_x = \omega_y = 2\pi\times 20$Hz, $ \omega_z = 2\pi\times 400$Hz.  The resultant interaction strengths for q2D $^{83}$Rb, $^{23}$Na, $^{87}$ Rb spin-2 BECs are
\begin{eqnarray*}
{(1)}\quad (c_0, c_1,c_2) &=& (392.14, -0.67, 2.74),\\
{(2)}\quad (c_0, c_1, c_2) &=& (136.18, 6.76, -7.30),\\
{(3)}\quad (c_0,c_1, c_2) &=& (465.92, 5.55, 0.18),
\end{eqnarray*}
respectively.
The oscillator lengths corresponding to three systems (1), (2) and (3) are $2.47~\mu$m, $4.69~\mu$m and $2.41~\mu$m, respectively. For these set of parameters, the ground state energies obtained with TSFS, TSBE, and TSCN are reported in Table {\ref{spin2-1d-eng}}. The agreement between the results with three methods is very good. 
\begin{table}[hbt!]
\addtolength{\tabcolsep}{-4pt}
\caption{Comparison of ground state energies of q1D and q2D spin-2 BECs of $^{83}$Rb, $^{23}$Na and  $^{87}$Rb obtained with TSFS, TSBE, and TSCN using $\Delta x =0.1$ and $\Delta {\tilde t}=0.001$ for different values of $\gamma$
and $\Omega$.}
\begin{center}
\begin{tabular}{c| c| c| c| c| c |c |c |c |c| c}
\hline
\multicolumn{1}{c}{}&
\multicolumn{1}{c}{}&
\multicolumn{3}{c}{$^{83}$Rb}&
\multicolumn{3}{c}{$^{23}$Na}&
\multicolumn{3}{c}{$^{87}$Rb}\\
\hline 
\multicolumn{1}{c}{}&
\multicolumn{1}{c}{$(\gamma,\Omega)$}&
\multicolumn{1}{ c }{TSFS}& 
\multicolumn{1}{ c } {TSBE}& 
\multicolumn{1}{ c }{TSCN}& 		
\multicolumn{1}{ c }{TSFS}& 
\multicolumn{1}{ c } {TSBE}& 
\multicolumn{1}{ c }{TSCN}& 		
\multicolumn{1}{ c }{TSFS}& 
\multicolumn{1}{ c } {TSBE}& 
\multicolumn{1}{ c }{TSCN}\\	
\hline
q1D& (0.5,0.5)&29.8496&29.8496&29.8496&14.6877 &14.6877 &14.6877 &34.2036&34.2036&34.2036\\ 
& (1.0,0.5)&28.3496&28.3499&28.3499&13.1877&13.1881&13.1881&32.7036&32.7039&32.7039\\ \hline
q2D& (0.5,0.1)&7.0875&7.0875&7.0875&3.9648&3.9645&3.9645&7.6850&7.6850&7.6850\\
\hline
\end{tabular}
\label{spin2-1d-eng}
\end{center}
\end{table}

\begin{figure}[hbt!]
\includegraphics[clip, trim=0.5cm 8cm 1.5cm 8cm,width=1.00\textwidth]{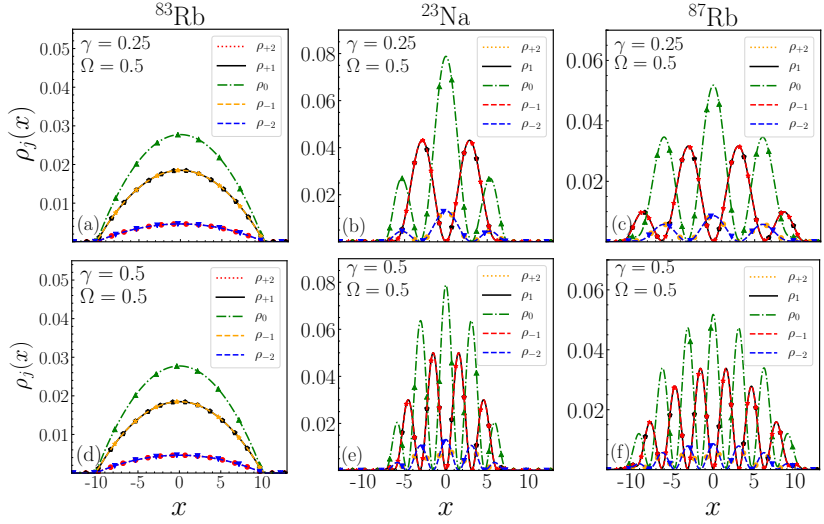}
\caption {(Color online) (a)-(c) are the ground-state 
component densities for SO-coupled quasi-1D $^{83}$Rb,
$^{23}$Na and $^{87}$Rb spin-2 BEC with $\gamma=0.25$ 
and $\Omega=0.5$ respectively, whereas (d)-(f) are the 
same  for $\gamma=0.5$ and $\Omega=0.5$. In (a)-(f), lines and 
points correspond to densities with TSBE and TSCN, respectively.}
\label{spin2+1d+den}
\end{figure}

\begin{figure}[H]
\begin{tabular}{c c c}
\includegraphics[width=12cm]{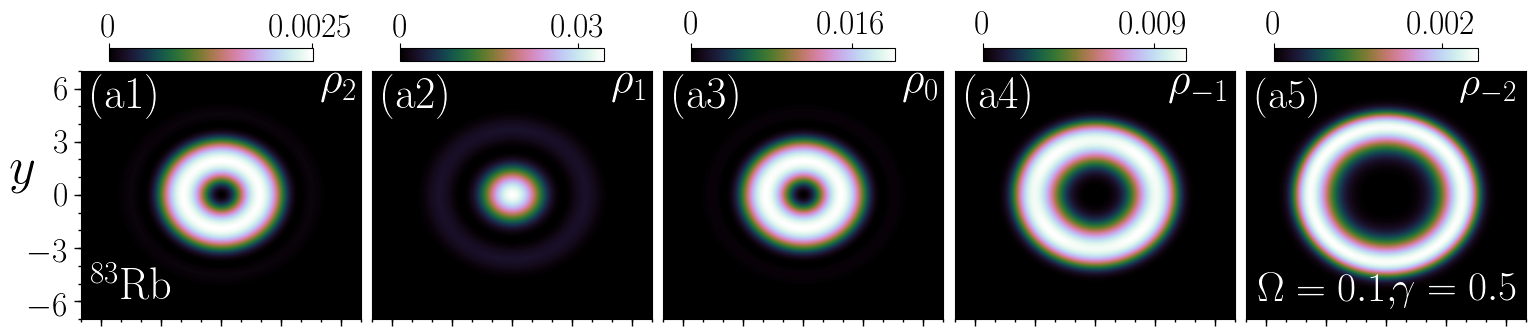}\\
\includegraphics[width=12cm]{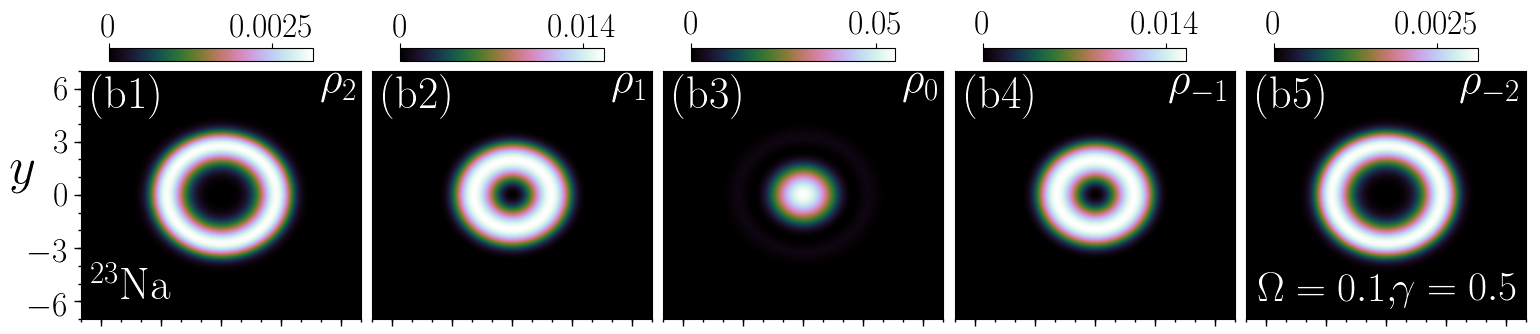}\\
\includegraphics[width=12cm]{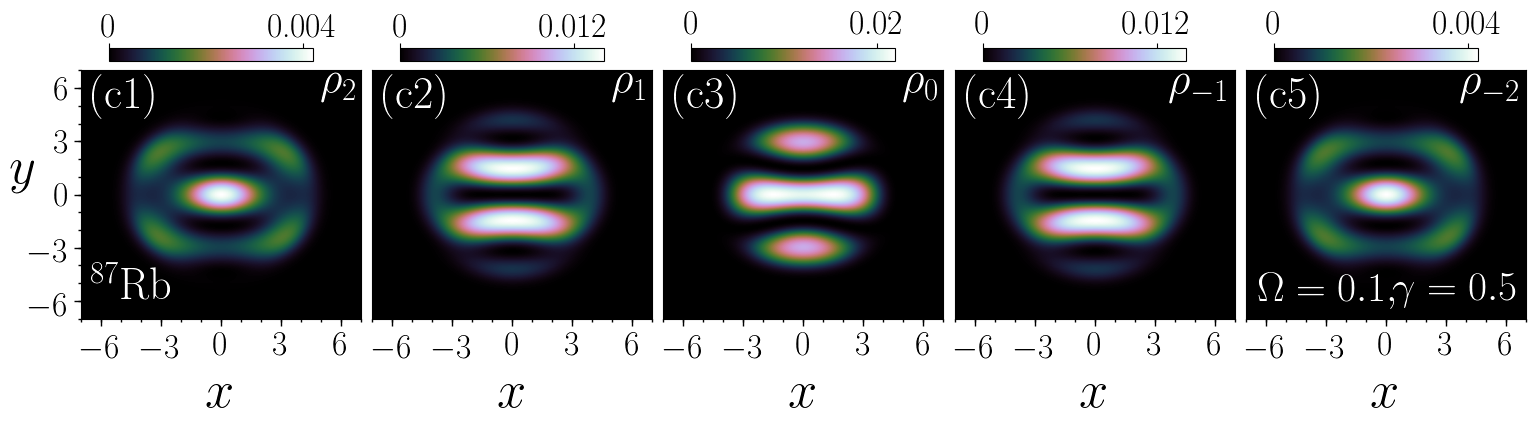}\\
\end{tabular}
\caption {(Color online) (a1)-(a5) are the component densities for an SO-coupled quasi-2D $^{83}$Rb,
(b1)-(b5) are for $^{23}$Na and (c1)-(c5) are for $^{87}$Rb spin-2 BEC with  $\Omega=0.1$, and $\gamma=0.5$
respectively.}
\label{spin2+2d+coh}
\end{figure}

Similarly, the ground-state component densities of q1D
spin-2 $^{83}$Rb, $^{23}$Na, and $^{87}$Rb BECs with 
different values of $\gamma$ and $\Omega$ calculated
using TSBE and TSCN are in very good agreement as shown in Fig. (\ref{spin2+1d+den}).
In q2D spin-2 BECs also, ground-state component densities 
calculated using three methods are in a very good agreement. 
Here we illustrate some qualitatively distinct ground-state
density profiles obtained with TSCN. The component densities 
in the ground state of q2D $^{83}$Rb, $^{23}$Na, and $^{87}$Rb
spin-2 BECs with $\gamma = 0.5$ and $\Omega =0.1$ are shown
in Fig.~\ref{spin2+2d+coh}. The ground state of $^{83}$Rb
and  $^{23}$Na spin-2 BECs have vortices of winding 
number $-1,0,+1,+2,+3$  and $-2,-1,0,+1,+2$ associated 
with the $l=2,1,0,-1,-2$ components, respectively. 
The ground state of q2D $^{87}$Rb spin-2 BEC has stripe
pattern in component densities for $\gamma=0.5$ and $\Omega=0.1$. 

\begin{figure}[H]
\begin{tabular}{c c}
\includegraphics[width=12cm]{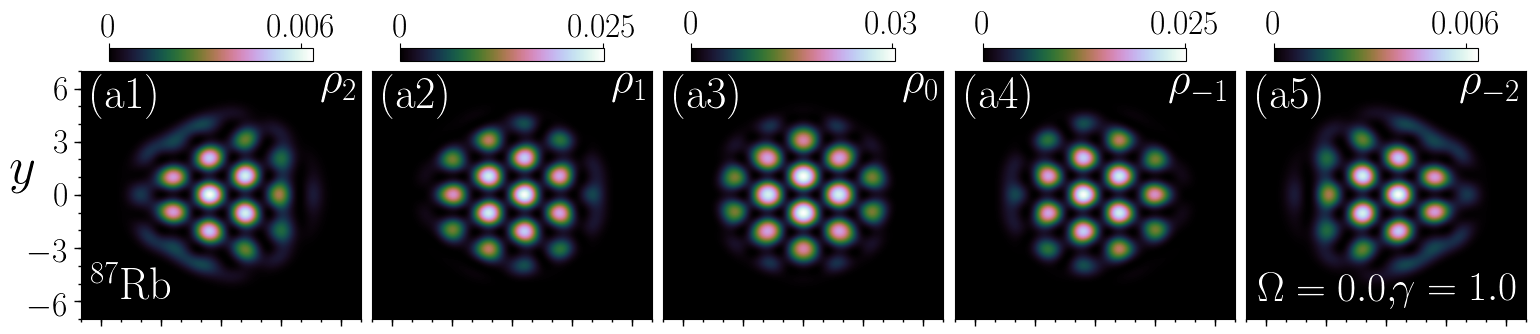}\\
\includegraphics[width=12cm]{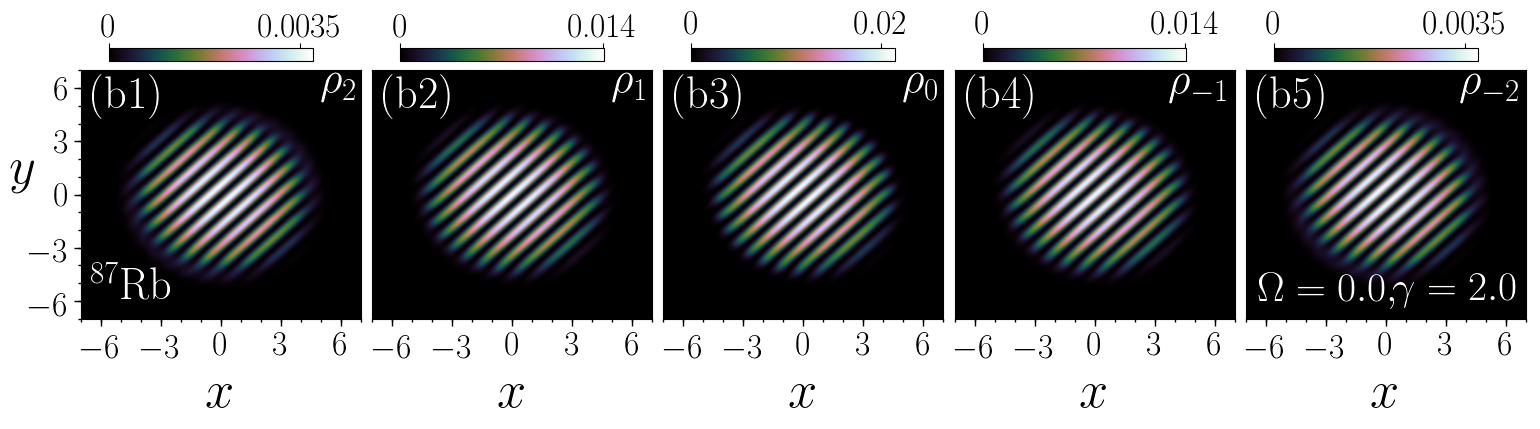}\\
\end{tabular}
\caption {(Color online) (a1)-(a5) are the component
densities for an SO-coupled quasi-2D $^{87}$Rb with 
$\gamma=1.0$ whereas (b1)-(b5) are for same system with $\gamma=2.0$ respectively.}
\label{spin2+2d+nocoh}
\end{figure}

For q2D $^{87}$Rb spin-2 BEC, the ground
state component densities with $\gamma =1, \Omega=0$ and
$\gamma=2,\Omega=0$ are also illustrated in Figs.~\ref{spin2+2d+nocoh}(a1)-(a5) and Figs.~\ref{spin2+2d+nocoh}(b1)-(b5), respectively. The ground-state component densities have triangular lattice
pattern for $\gamma=1$ and stripe density pattern for $\gamma=2$.

\section*{Summary}
We have discussed time-splitting Backward-Euler and 
Crank-Nicolson methods to study the SO-coupled spinor BECs 
with coherent coupling. 
We have developed the methods for pseudospin-1/2, spin-1 and
spin-2 BEC in q1D and q2D traps. We have considered Rashba SO
coupling in the present work, one can also consider Dresselhaus
coupling or a combination of both within the framework of same numerical schemes.
We have compared the results obtained with these finite
difference methods 
with the time-splitting Fourier spectral method. 
The numerical results for stationary states
obtained with the three methods are in very good 
agreement. We have provided the comparison of ground state
energies and component density profiles calculated using
three methods for several illustrative cases. 
In imaginary-time propagation, TSCN shows faster
convergence as compared to TSBE. Moreover, the time 
evolution as per TSCN is unitary time evolution consistent
with the underlying Hermitian Hamiltonian. This is not the
case with TSBE which results in non-unitary time evolution
and thus rendering the method not suitable to the study any
real-time dynamics. The finite difference methods developed
in the present work can be easily extended to higher spin system like spin-3 BEC.
\section*{Appendix}
\label{appendix1}
The split equation for ${H}_{\rm nd+}$ is
\begin{equation}
i\frac{\partial { \Psi}}{\partial t} = {H}_{\rm nd+}{\Psi}
\label{App_nd_spin2}
\end{equation}
where
\begin{equation}
H_{\rm nd+} = 
\begin{pmatrix}
0 & h_{12} &h_{13}  & 0 &0\\
h_{12}^* &0 &h_{23}  & 0 &0\\
h_{13}^* & h_{23}^* & 0 & h_{34} &h_{35} \\
0 &0  & h_{34}^* &0 &h_{45} \\
0 &0  &h_{35}^* & h_{45}^*  & 0
\end{pmatrix}{\label{SEmatrix}},
\end{equation}
Eq. (\ref{SEmatrix}) can split into two operator $H_{\rm{1}}$ and $H_{\rm{2}}$
\begin{equation}
H_{\rm 1} = 
\begin{pmatrix}
0 & h_{12} &0  & 0 &0\\
h_{12}^* &0 &h_{23}  & 0 &0\\
0 & h_{23}^* & 0 & h_{34} &0 \\
0 &0  & h_{34}^* &0 &h_{45} \\
0 &0  &0& h_{45}^*  & 0
\end{pmatrix},\quad
H_{\rm 2} = 
\begin{pmatrix}
0 & 0 &h_{13}  & 0 &0\\
0 &0 &0  & 0 &0\\
h_{13}^* & 0 & 0 & 0 &h_{35} \\
0 &0  & 0 &0 &0 \\
0 &0  &h_{35}^* & 0  & 0
\end{pmatrix},\\
\end{equation}
where
\begin{align*}
h_{12} &=  c_1 F_{-} - \frac{2}{5}c_2\psi_{-1} \psi_{-2}^* +\frac{\Omega}{2},\quad
h_{23} =  \frac{\sqrt{6}}{2}c_1 F_{-}- \frac{1}{5}
c_2\psi_{0} \psi_{-1}^* +\frac{\sqrt{6}\Omega}{4},\\
h_{34} &= \frac{\sqrt{6}}{2}c_1 F_{-}- \frac{1}{5}c_2\psi_{1} 
\psi_{0}^* +\frac{\sqrt{6}\Omega}{4},\quad h_{45} =  c_1 F_{-} - \frac{2}{5} c_2\psi_{2}
\psi_{1}^* +\frac{\Omega}{2},\\
h_{13} &= \frac{1}{5}c_2\psi_{0} \psi_{-2}^*, \quad
h_{35} = \frac{1}{5}c_2\psi_{2} \psi_{0}^*.\\
\end{align*}
The approximate solution of Eq.~(\ref{App_nd_spin2}) is given by
\begin{eqnarray}
\Psi(x,t +\delta t) &=&
\exp\left(-iH_{\rm nd+}dt\right)\Psi(x,t),\nonumber\\
 &\approx&
\exp\left(-iH_{\rm 2}dt\right)\exp\left(-iH_{\rm 1}dt\right)\Psi(x,t),\nonumber\\
               &=& \exp(-i\delta t P A_2 P^{-1}) \exp(-i\delta t S A_1 S^{-1})\Psi(x,t),\nonumber\\
             &=& P\exp(-i\delta t A_2 ) P^{-1}S\exp(-i\delta t A_1 ) S^{-1} \Psi(x,t),
             \label{H1_nd_solu}
\end{eqnarray}
where $5\times 5$ matrix
\begin{equation}
S = \begin{pmatrix}
u_1, &u_2, &u_3,& u_4& u_5\end{pmatrix}.\nonumber\\
\end{equation}
The ($u_1,u_2,u_3,u_4,u_5$) are normalised eigen vectors which can be obtained from un-normalised eigen vectors ($v_1,v_2,v_3,v_4,v_5$), defined as  
\begin{eqnarray}
v_1 &=&\Bigg[\frac{h_{23}h_{45}}{h_{12}^* h_{34}^*},0,-\frac{h_{45}}{h_{34}^*},0,1\Bigg]^T,\nonumber\\
v_2 &=& \Bigg[\frac{h_{12} \left(-\alpha ^2+\beta ^2+2 |h_{12}| ^2+2 | h_{23}| ^2-2 |h_{34}| ^2-2|
h_{45}| ^2\right)}{4 h_{23}^* h_{34}^* h_{45}^*},\nonumber\\ && \frac{ \beta \left(\alpha ^2-\beta^2
+2 |h_{12}| ^2-2 | h_{23}| ^2+2 |h_{34}| ^2+2| h_{45}| ^2\right)}{4\sqrt{2} h_{23}^* h_{34}^* h_{45}^*}
,\nonumber\\ && \frac{\left(-\alpha ^2+\beta ^2+2 |h_{12}| ^2+2 | h_{23}| ^2+2 |h_{34}| ^2-2| h_{45}|^2
\right)}{4  h_{34}^* h_{45}^*}, -\frac{\beta }{\sqrt{2} h_{45}^*},1 \Bigg]^T,\nonumber\\
v_3 &=& \Bigg[\frac{h_{12} \left(-\alpha ^2+\beta ^2+2 |h_{12}| ^2+2 | h_{23}| ^2-2 |h_{34}| ^2
-2| h_{45}| ^2\right)}{4 h_{23}^* h_{34}^* h_{45}^*},\nonumber\\ &&- \frac{ \beta \left(\alpha^2
-\beta ^2+2 |h_{12}| ^2-2 | h_{23}| ^2+2 |h_{34}| ^2+2| h_{45}| ^2\right)}{4\sqrt{2} h_{23}^*
h_{34}^* h_{45}^*},\nonumber\\ && \frac{\left(-\alpha ^2+\beta ^2+2 |h_{12}| ^2+2 | h_{23}|^2
+2 |h_{34}| ^2-2| h_{45}| ^2\right)}{4  h_{34}^* h_{45}^*}, \frac{\beta }{\sqrt{2} h_{45}^*},
1 \Bigg]^T,\nonumber
\end{eqnarray}
\begin{eqnarray}
v_4 &=& \Bigg[\frac{h_{12} \left(\alpha ^2-\beta ^2+2 |h_{12}| ^2+2 | h_{23}| ^2-2 |h_{34}| ^2
-2| h_{45}| ^2\right)}{4 h_{23}^* h_{34}^* h_{45}^*},\nonumber\\ && \frac{ \alpha \left(-\alpha^2
+\beta ^2+2 |h_{12}| ^2-2 | h_{23}| ^2+2 |h_{34}| ^2+2| h_{45}| ^2\right)}{4\sqrt{2} h_{23}^* h_{34}^*
h_{45}^*},\nonumber\\ && \frac{\left(\alpha ^2-\beta ^2+2 |h_{12}| ^2+2 | h_{23}| ^2+2 |h_{34}| ^2-2|
h_{45}| ^2\right)}{4  h_{34}^* h_{45}^*}, -\frac{\alpha }{\sqrt{2} h_{45}^*},1 \Bigg]^T,\nonumber\\
v_5 &=& \Bigg[\frac{h_{12} \left(\alpha ^2-\beta ^2+2 |h_{12}| ^2+2 | h_{23}| ^2-2 |h_{34}| ^2-2|
h_{45}| ^2\right)}{4 h_{23}^* h_{34}^* h_{45}^*},\nonumber\\ && -\frac{ \alpha \left(-\alpha ^2+
\beta ^2+2 |h_{12}| ^2-2 | h_{23}| ^2+2 |h_{34}| ^2+2| h_{45}| ^2\right)}{4\sqrt{2} h_{23}^* 
h_{34}^* h_{45}^*},\nonumber\\ && \frac{\left(\alpha ^2-\beta ^2+2 |h_{12}| ^2+2 | h_{23}| ^2+
2 |h_{34}| ^2-2| h_{45}| ^2\right)}{4  h_{34}^* h_{45}^*}, \frac{\alpha }{\sqrt{2} h_{45}^*},1 \Bigg]^T,
\label{vector_h1}
\end{eqnarray}
 by using Gram-Schmidt orthogonalization. The matrix
\begin{equation}
A_1 =\text{diag} \left( 0,-\frac{\beta }{\sqrt{2}},\frac{\beta }{\sqrt{2}},-\frac{\alpha }{\sqrt{2}},
\frac{\alpha }{\sqrt{2}} \right),
\end{equation}
where 
\begin{eqnarray}
\alpha ^2 &=& \sqrt{(|h_{12}|^2 + |h_{23}|^2 + |h_{34}|^2 + |h_{45}|^2)^2-4(|h_{12}|^2|h_{34}|^2 +  
|h_{45}|^2(|h_{12}|^2 + |h_{23}|^2))}\nonumber\\ && +|h_{12}|^2 + |h_{23}|^2 + |h_{34}|^2 + 
|h_{45}|^2,\nonumber\\
\beta ^2 &=& - \sqrt{(|h_{12}|^2 + |h_{23}|^2 + |h_{34}|^2 + |h_{45}|^2)^2-4(|h_{12}|^2|h_{34}|^2 
+ |h_{45}|^2(|h_{12}|^2 + |h_{23}|^2))}\nonumber\\ && +|h_{12}|^2 + |h_{23}|^2 + |h_{34}|^2 +
|h_{45}|^2.\nonumber\\
\nonumber
\end{eqnarray}
Similarly, $5\times5$ matrix 
\begin{equation}
P = \begin{pmatrix}
w_1, &w_2, &w_3,& w_4& w_5\end{pmatrix},\nonumber
\end{equation}
where 
\begin{eqnarray}
w_1 &=& \Bigg[ -\frac{h_{35} |h_{13}|}{h_{13}^* \sqrt{|h_{13}| ^2+|h_{35}| ^2}},0,0,0,\frac{1}
{\sqrt{\frac{h_{35} h_{35}^*}{|h_{13}| ^2}+1}}\bigg]^T,\nonumber\\ w_2 &=& [0, 0, 0, 1, 0]^T,
\quad w_3 =[0, 1, 0, 0, 0]^T,\nonumber\\ w_4 &=& \Bigg[ \frac{h_{13}|h_{35}|}{\sqrt{2}h_{35}^*
\sqrt{|h_{13}| ^2+|h_{35}| ^2}},0,-\frac{|h_{35}|}{\sqrt{2}h_{35}^*},0,\frac{|h_{35}|}{\sqrt{2}
\sqrt{|h_{13}|^2+|h_{35}| ^2}} \bigg]^T,\nonumber\\ w_5 &=& \Bigg[ \frac{h_{13}|h_{35}|}
{\sqrt{2}h_{35}^* \sqrt{|h_{13}| ^2+|h_{35}| ^2}},0,\frac{|h_{35}|}{\sqrt{2}h_{35}^*},
0,\frac{|h_{35}|}{\sqrt{2} \sqrt{|h_{13}|^2+|h_{35}| ^2}} \bigg]^T,
\end{eqnarray}
and $5\times5$ matrix
\begin{equation}
A_2 = \text{diag}\left(0,0,0,- \sqrt{|h_{13}|^2+|h_{35}| ^2}, \sqrt{|h_{13}|^2+|h_{35}| ^2}\right).
\end{equation}

\end{document}